\documentclass[11pt]{article}

\usepackage[preprint]{acl}

\usepackage{times}
\usepackage{latexsym}
\usepackage{amsmath}
\usepackage{amssymb}
\usepackage{algorithm}
\usepackage{booktabs}
\usepackage{longtable}
\usepackage{mdframed}
\usepackage{multirow}
\usepackage{wrapfig}

\usepackage{algorithmicx} \usepackage{algpseudocode} 

\usepackage[T1]{fontenc}

\usepackage[utf8]{inputenc}

\usepackage{microtype}

\usepackage{inconsolata}

\usepackage{graphicx}

\title{IRAH: Hierarchical Agentic Tree Search for Retrieval Pipeline Optimization}

\author{
Maria Ana Cardei$^{1,\dagger}$ \thanks{cbr8ru@virginia.edu},
Aniket Singh$^{2,\dagger}$,
Amir Moghadam$^{3}$,
Srividhya Baikadi$^{3}$,
Rashmi Shetty$^{3}$ \\
$^{1}$University of Virginia \quad
$^{2}$University of California San Diego \quad
$^{3}$Capital One \\
$^{\dagger}$Equal contribution; work conducted while both authors were interns at Capital One.
}

\begin{document}
\maketitle
\begin{abstract}
Optimizing Information Retrieval (IR) pipelines for Retrieval-Augmented
Generation (RAG) requires jointly selecting structural components and
numerical hyperparameters under limited evaluation budgets. We introduce the
\textbf{I}nformation \textbf{R}etrieval \textbf{A}gentic
\textbf{H}arness (\mbox{IRAH}), which combines Language Agent Tree Search
for IR (\mbox{LATS-IR}) with offline meta-learned initialization.
\mbox{LATS-IR} performs UCT-guided hierarchical search, first exploring
macro-structural pipeline choices and then refining numerical parameters,
with natural-language reflections to guide subsequent mutations and provide
inspectable search traces. We evaluate \mbox{IRAH} on three in-distribution benchmarks and one out-of-distribution benchmark to assess generalization under distribution shift. Under the same meta-learned initialization,
\mbox{LATS-IR} outperforms TPE on two of three in-distribution benchmarks, with gains of up to $6.4\%$, and remains competitive on the out-of-distribution benchmark.
Across all four benchmarks, warm-started
\mbox{IRAH} achieves the highest average held-out $\mathrm{nDCG}@10$
($0.4816$) and improves over its cold-start counterpart.
These results establish
\mbox{IRAH} as a competitive approach to retrieval pipeline optimization
under constrained evaluation budgets.
\end{abstract}

\section{Introduction}

Modern retrieval-augmented generation (RAG) systems rely on retrieval
pipelines whose effectiveness is shaped by interdependent design choices
\citep{lewis2020retrieval,hsia2025ragged}. These include how documents are
chunked, which dense encoder is used, whether sparse and dense scores are
fused, which reranker is applied, and how numerical parameters such as fusion
weight and retrieval depth are set. A chunking strategy changes the
distribution of candidate passages seen by the embedder; reranking can
compensate for some retrieval errors while amplifying others; and the best
retrieval depth depends on both the underlying corpus and the evaluation
objective \citep{hsia2025ragged}. As a result, RAG pipeline optimization is
better viewed as a structured configuration-search problem than as isolated
tuning of individual hyperparameters.

Existing automated optimization methods reduce manual tuning burden
\citep{akiba2019optuna}, but general-purpose black-box optimization does not
explicitly exploit the macro--micro structure of retrieval pipeline decisions.
Recent work on RAG hyperparameter optimization demonstrates the value of
structured search in large configuration spaces
\citep{fu-etal-2024-autorag}. However, existing approaches do not explicitly
organize optimization as a hierarchical search trajectory in which structural
and numerical decisions are explored at different stages and prior outcomes
inform subsequent proposals. Moreover, many hyperparameter optimization
algorithms provide limited visibility into the optimization process
\citep{moosbauer2022improving}. For IR pipelines, inefficient exploration can
also incur substantial systems costs when structural changes require
constructing or loading new indices. Starting optimization from scratch for
each new corpus further discards potentially useful information from prior
retrieval tasks. These limitations motivate an approach that can
exploit the structure of the configuration space, adapt its search based on
previous evaluations, transfer useful experience across datasets, and retain
an inspectable optimization trace.

\begin{figure*}[t]
    \centering
    \includegraphics[width=0.9\textwidth]{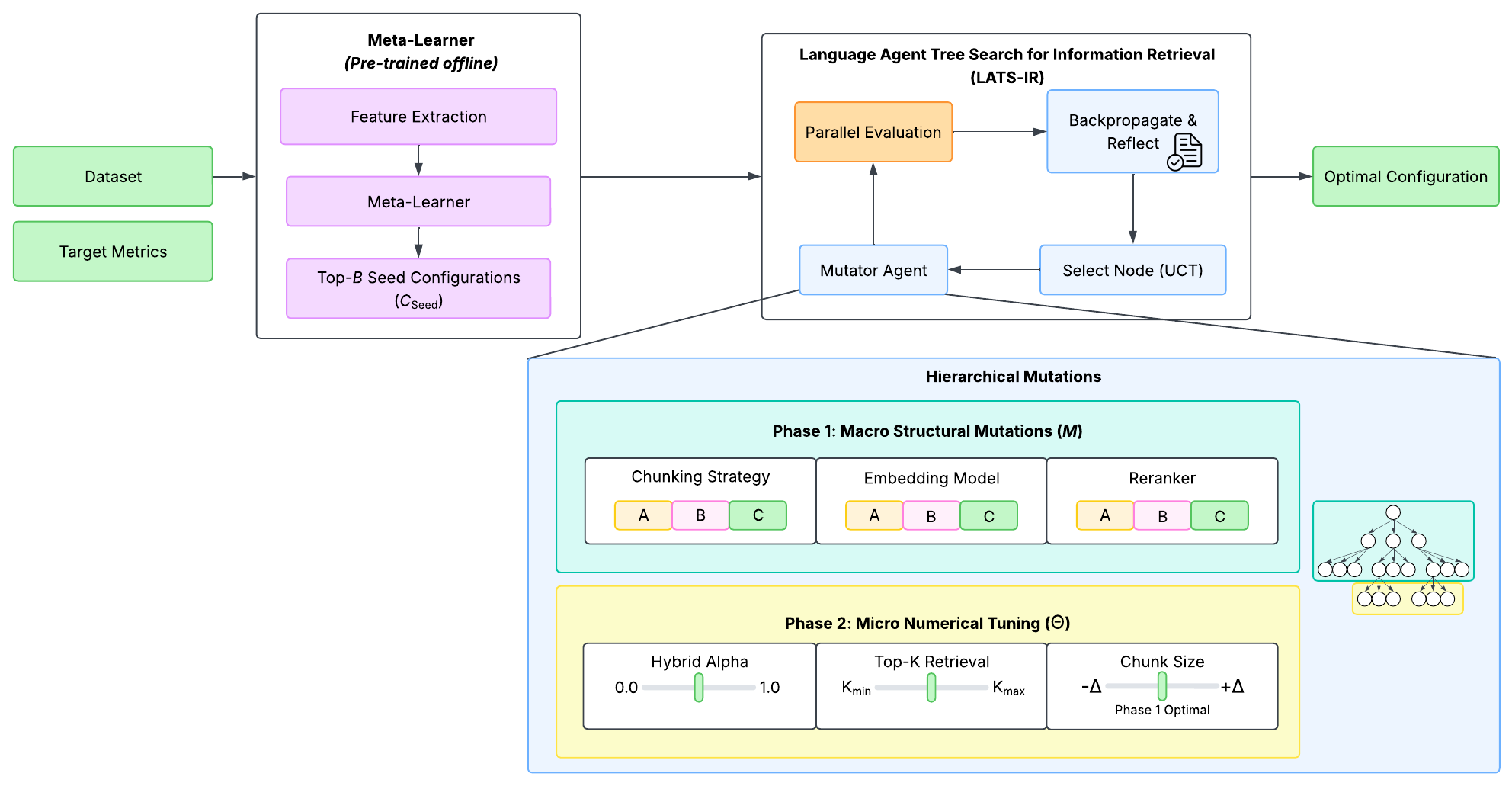}
    \caption{\mbox{\textbf{IRAH}}: Given a target dataset and task metric, an
    offline meta-learner proposes promising seed configurations.
    LATS-IR then performs two-phase hierarchical search over structural and
    numerical pipeline decisions and returns the best configuration identified under the evaluation budget.}
    \label{fig:architecture}
    \vspace{-14pt}
\end{figure*}

We introduce the \textbf{I}nformation \textbf{R}etrieval \textbf{A}gentic
\textbf{H}arness \textbf{(\mbox{IRAH})} (Figure~\ref{fig:architecture}),
which combines \mbox{LATS-IR}, a reflection-guided hierarchical search
process, with offline meta-learned initialization. LATS-IR first performs macro-structural exploration, evaluating structural
choices together with coarse numerical settings, before locally refining
numerical parameters within a promising pipeline architecture. Upper Confidence Bound applied to Trees (UCT) guides tree
selection, while natural-language reflections from evaluated configurations
condition subsequent mutations and provide an inspectable record of the
search process. To mitigate cold-start overhead, an offline meta-learner uses
dataset-level characteristics to identify promising configurations from prior
retrieval tasks for warm-start initialization.

We evaluate \mbox{IRAH} on three in-distribution retrieval benchmarks and one
out-of-distribution benchmarks under a fixed evaluation budget. Under the same
meta-learned initialization, \mbox{LATS-IR} improves held-out
$\mathrm{nDCG}@10$ over TPE by up to $6.4\%$ on the in-distribution benchmarks
and remains competitive on the out-of-distribution target. Across all four benchmarks, warm-started \mbox{IRAH} achieves the highest average
held-out $\mathrm{nDCG}@10$ and outperforms its cold-start counterpart.
These results demonstrate the effectiveness of combining hierarchical agentic search with meta-learned initialization for retrieval
pipeline optimization under constrained evaluation budgets.

Our main contributions are as follows:
\begin{enumerate}
    \item We introduce \mbox{LATS-IR}, a reflection-guided hierarchical
    search method for IR pipeline optimization that separates macro-structural
    exploration from local numerical refinement and uses prior evaluations
    to guide subsequent search.
    \item We integrate \mbox{LATS-IR} with offline meta-learning in
    \mbox{IRAH}, allowing promising configurations from prior retrieval tasks
    to warm-start hierarchical search.
    \item We evaluate \mbox{IRAH} across three in-distribution and one
    out-of-distribution retrieval benchmarks. Under shared initialization,
    \mbox{LATS-IR} outperforms TPE on two of three in-distribution benchmarks.
    Warm-starting improves \mbox{IRAH} on all four benchmarks, and
    performance remains competitive on the out-of-distribution target.
\end{enumerate}

\section{Related Work}

\paragraph{Retrieval-Augmented Generation and Pipeline Design.}
Retrieval-augmented generation (RAG) grounds generation in externally
retrieved evidence \citep{lewis2020retrieval}, with performance depending on
interacting choices in preprocessing, chunking, retrieval, reranking, and
generation \citep{gao2023rag}. Systems such as DSPy treat language-model
applications as compositional pipelines whose components can be optimized
\citep{khattab2024dspy}. We focus on retrieval configuration, where structural
and numerical choices jointly affect effectiveness and systems costs such as
index construction and reuse.

\paragraph{Hyperparameter and Structured RAG Optimization.}
General-purpose methods such as Tree-structured Parzen Estimator (TPE)
\citep{bergstra2011algorithms,akiba2019optuna} and AutoML under the CASH
formulation \citep{thornton2013auto,feurer2015efficient} automate configuration
search from observed evaluations, but do not explicitly encode the
retrieval-specific hierarchy considered here.

Recent RAG optimization methods increasingly exploit search-space structure.
\citet{fu-etal-2024-autorag} use a two-level hierarchical bandit;
\citet{afzal-etal-2026-beyond-grid} compare global and component-wise Bayesian
optimization; and \citet{ijcai2026p466} use cyclic sequential optimization of
retriever and generator parameters. Language models have also been used as
iterative optimizers that condition proposals on prior outcomes
\citep{yang2024large}. In contrast, \mbox{LATS-IR} organizes structural and
coarse numerical exploration separately from local numerical refinement while
maintaining an explicit search tree over retrieval configurations.

\paragraph{Tree Search and Language-Agent Reflection.}
Monte Carlo Tree Search and UCT balance exploitation and exploration
\citep{kocsis2006bandit,browne2012survey}. Language Agent Tree Search (LATS)
extends tree search to language-model agents using generation and
self-reflection \citep{zhou2023language}. \mbox{LATS-IR} adapts these ideas to
retrieval configuration optimization: nodes are evaluated pipeline
configurations, edges are phase-constrained mutations, UCT guides selection,
and reflections condition subsequent candidate generation.

\paragraph{Meta-Learning for Warm Starts.}
Meta-learning transfers information across tasks to improve learning or
optimization \citep{vanschoren2019meta}. AutoML systems use dataset
meta-features and prior evaluations to identify promising initial
configurations \citep{feurer2015efficient}. \mbox{IRAH} applies this idea to
retrieval optimization by using an offline meta-learner to rank shared anchor
configurations and supply the highest-ranked candidates as warm starts for
\mbox{LATS-IR}.

\section{IRAH: Information Retrieval Agentic Harness}

In this section, we introduce \mbox{IRAH}, a framework for retrieval
configuration optimization. \mbox{IRAH} centers on \mbox{LATS-IR}, a
reflection-guided hierarchical search procedure, and uses offline meta-learning
to provide warm-start initialization. We formalize the hierarchical configuration problem (Section \ref{subsec:prelim}),
then describe \mbox{LATS-IR} (Section \ref{subsec:lats_reflection}), and finally introduce the meta-learned
warm-start mechanism (Section \ref{subsec:meta_learner}).

\subsection{Preliminaries} \label{subsec:prelim}

We formulate information retrieval (IR) configuration optimization as a
hierarchical configuration-search problem, extending the Combined Algorithm
Selection and Hyperparameter Optimization (CASH) paradigm
\citep{thornton2013auto,feurer2015efficient} to modular retrieval pipelines
used in RAG systems \citep{gao2023rag,khattab2024dspy}. Rather than treating
all pipeline decisions as an undifferentiated set of parameters, we distinguish
between \emph{macro-structural} decisions and \emph{micro-level} numerical
parameters. A candidate configuration is
$\mathbf{c}=(\mathbf{m},\boldsymbol{\theta})$, where
$\mathbf{m}\in\mathcal{M}$ denotes categorical structural choices, such as
the chunking strategy or embedding model, and
$\boldsymbol{\theta}\in\Theta(\mathbf{m})$ denotes numerical parameters
associated with that structure, such as hybrid fusion weight $\alpha$.

This hierarchy reflects both the dependencies and computational costs of
retrieval pipeline optimization. Numerical settings are tuned within the
context of a particular pipeline structure, while changing structural choices
such as the chunking strategy or embedding model can require constructing a
new physical index. In contrast, parameters such as retrieval depth or fusion
weight can often be varied while reusing an existing index. Prior RAG
optimization work similarly demonstrates benefits from structured or
component-wise search
\citep{fu-etal-2024-autorag,afzal-etal-2026-beyond-grid,ijcai2026p466}.
We therefore define the configuration space as
\[
\mathcal{S}
=
\left\{
(\mathbf{m},\boldsymbol{\theta})
\;:\;
\mathbf{m}\in\mathcal{M},
\;
\boldsymbol{\theta}\in\Theta(\mathbf{m})
\right\}.
\]
Given a target dataset and an optimization-query set
$Q_{\mathrm{opt}}$, the objective is to identify the configuration that
maximizes retrieval effectiveness:
\begin{equation}
\mathbf{c}^{*}
=
\arg\max_{\mathbf{c}\in\mathcal{S}}
\frac{1}{|Q_{\mathrm{opt}}|}
\sum_{q\in Q_{\mathrm{opt}}}
R(\mathbf{c};q),
\end{equation}
where $R(\mathbf{c};q)$ denotes the retrieval utility of configuration
$\mathbf{c}$ for query $q$, measured using a task-specific metric such as
$\mathrm{nDCG}@k$ or $\mathrm{Recall}@k$.

Because evaluating configurations can be computationally expensive, we
consider a budgeted setting in which the optimizer may evaluate at most
$N_{\max}$ configurations. The goal is therefore to identify a
high-performing approximation to $\mathbf{c}^{*}$ within a limited
evaluation budget. This budgeted setting motivates allocating evaluations strategically between
structural exploration and numerical refinement.

\begin{algorithm}[t]
\caption{\mbox{LATS-IR} Hierarchical Retrieval Configuration Search}
\label{alg:latsir}
\begin{algorithmic}[1]
\Require optional seeds $\mathcal{C}_{\mathrm{seed}}$,
         evaluation budget $N_{\max}$,
         Phase~1 budget $N_1$,
         candidates per expansion $K$

\State $\mathcal{T},\mathcal{H},n
       \gets \Call{Initialize}
       {\mathcal{C}_{\mathrm{seed}},Q_{\mathrm{opt}}}$

\While{$n < N_{\max}$}

    \If{$n \geq N_1$ and Phase~2 not initialized}
        \State $\mathbf{c}^{(1)*}
        \gets \Call{BestPhase1}{\mathcal{T}}$
    \EndIf

    \State $\mathcal{S} \gets
    \begin{cases}
        \mathcal{S}_{\mathrm{P1}}, & n < N_1,\\
        \Theta_{\mathrm{local}}(\mathbf{c}^{(1)*}), & n \geq N_1
    \end{cases}$

    \State $w \gets \Call{AdaptiveUCT}{\mathcal{T},n}$
    \State $\mathbf{c}_p \gets \Call{Select}{\mathcal{T},w}$

    \State $\mathcal{C}
        \gets \Call{Mutate}
        {\mathbf{c}_p,\mathcal{H},\mathcal{S},K}$
    \State $\mathcal{C}
        \gets \Call{Validate}{\mathcal{C},\mathcal{S}}$

    \ForAll{$\mathbf{c}\in\mathcal{C}$}
        \State $r \gets \Call{Evaluate}{\mathbf{c},Q_{\mathrm{opt}}}$
        \State $h \gets \Call{Reflect}{\mathbf{c},\mathbf{c}_p,r}$
        \State $\mathcal{H}\gets\mathcal{H}\cup\{h\}$
        \State $\Call{AddChild}
               {\mathcal{T},\mathbf{c}_p,\mathbf{c},r,h}$
        \State $\Call{Backpropagate}{\mathcal{T},\mathbf{c},r}$
        \State $n\gets n+1$
    \EndFor
\EndWhile

\State \Return $\Call{BestEvaluated}{\mathcal{T}}$
\end{algorithmic}
\end{algorithm}

\subsection{LATS-IR: Language Agent Tree Search for Information Retrieval}
\label{subsec:lats_reflection}

\mbox{LATS-IR} is our optimization engine. It adapts Language Agent Tree Search
(LATS) \citep{zhou2023language} to the hierarchical IR configuration problem
defined in Section~\ref{subsec:prelim}. The hierarchy is imposed through the
search procedure: tree nodes represent fully specified retrieval configurations,
while phase-constrained edges first permit macro-structural changes and later
restrict search to local numerical refinement. \mbox{LATS-IR} can operate from an uninformed cold start with agent-generated
initial configurations, or from configurations supplied by the meta-learned
warm-start mechanism described in Section~\ref{subsec:meta_learner}. For a
warm start, the supplied seeds are evaluated and the highest-reward seed
initializes the tree root.

At each tree expansion, \mbox{LATS-IR} (1) selects a previously evaluated
node, (2) generates $K$ valid candidate mutations conditioned on the active
search phase, and (3) evaluates the candidates on $Q_{\mathrm{opt}}$.
The candidates are added as sibling children of the selected parent, and each
reward is backpropagated through its tree path to update visit counts and
empirical value estimates. This process repeats until the evaluation budget
$N_{\max}$ is exhausted. Algorithm~\ref{alg:latsir} summarizes \mbox{LATS-IR}.

\paragraph{UCT-Guided Node Selection.}
Each evaluated configuration produces a retrieval reward $r_t$, which is
backpropagated along the selected tree path. For each node on this path, we
increment its visit count and update its empirical value
$\bar{V}(\mathbf{c})$ as the mean reward backpropagated through that node.
Consequently, ancestor nodes summarize the observed performance of
configurations explored within their descendant branches.

Nodes are selected for expansion using Upper Confidence Bounds for Trees
(UCT) \citep{kocsis2006bandit}:
\begin{equation}
\label{eq:uct}
\mathrm{UCT}(\mathbf{c})
=
\bar{V}(\mathbf{c})
+
w_t
\sqrt{
\frac{\ln N(\mathbf{c}_{\mathrm{parent}})}
{N(\mathbf{c})}
},
\end{equation}
where $N(\mathbf{c})$ and $N(\mathbf{c}_{\mathrm{parent}})$ are the visit
counts of the node and its parent, respectively, and $w_t$ controls the
exploration-exploitation trade-off. Rather than using a fixed exploration
coefficient, \mbox{LATS-IR} adapts $w_t$ to the current search phase and
observed reward history. Phase~1 favors broader exploration, whereas Phase~2
places greater emphasis on local exploitation. The complete adaptive schedule
is provided in Appendix~\ref{app:adaptive_uct}.

\paragraph{Two-Phase Hierarchical Search.}
LATS-IR divides optimization into a macro phase centered on structural
exploration followed by micro-level numerical refinement, exploiting the
structure of the configuration space $\mathcal{S}$ under a limited evaluation
budget.
In \textbf{Phase~1}, candidate mutations may modify structural components
$\mathbf{m}$, including the embedding model, chunking strategy, and reranker,
together with coarse numerical settings. The goal is to explore alternative
retrieval architectures and identify a promising structural region of the
search space.
At the phase transition, we select the Phase~1 configuration with the highest
optimization reward,
$\mathbf{c}^{(1)*}=(\mathbf{m}^{*},\boldsymbol{\theta}^{(1)*})$,
and fix its categorical structural choices $\mathbf{m}^{*}$.
\textbf{Phase~2} then refines the numerical parameters within a local
neighborhood
$\Theta_{\mathrm{local}}(\mathbf{c}^{(1)*})
\subseteq \Theta(\mathbf{m}^{*})$
centered on the Phase~1 configuration.
This separation prevents the search from revisiting alternative
macro-architectures during refinement and concentrates the remaining budget
on fine-grained optimization of a promising structural configuration.

\paragraph{Reflection-Guided Candidate Generation.}

Candidate mutations are generated by an LLM conditioned on the selected
parent configuration, its observed reward, relevant prior reflections, the
active search phase, and the valid parameter choices or numerical bounds for
that phase. For each expansion, the LLM proposes $K$ structured child
configurations, which are validated against the allowable search space before
evaluation.
After each evaluation, the agent produces a short reflection comparing the
evaluated configuration with its parent and, when available, their observed
reward difference. The reflection is retained in search memory and supplied
as context for subsequent candidate generation, allowing later mutations to
incorporate information from earlier successes and failures. Complete
prompting, validation, and output-schema details are provided in
Appendix~\ref{app:reflection_mutation}.

\subsection{Meta-Learned Warm-Start Initialization}
\label{subsec:meta_learner}

To mitigate cold-start overhead, we construct an offline meta-learner
$f_{\mathrm{meta}}$ from configuration evaluations collected across multiple
source retrieval datasets, following meta-learning approaches in AutoML
\citep{feurer2015efficient,vanschoren2019meta}.
For an IR dataset
$\mathcal{D}=(\mathcal{C},\mathcal{Q})$, comprising corpus $\mathcal{C}$
and query set $\mathcal{Q}$, we extract a dataset-level representation
$\phi(\mathcal{D}) \in \mathbb{R}^{d}$ capturing corpus, query, and
query-corpus characteristics. Each candidate retrieval configuration
$\mathbf{c}$ is represented by configuration features
$\psi(\mathbf{c}) \in \mathbb{R}^{p}$.

Given a dataset-configuration pair, the meta-learner produces a score
indicating the predicted utility of that configuration for the dataset:
\begin{equation}
s_{\mathrm{meta}}(\mathbf{c},\mathcal{D})
=
f_{\mathrm{meta}}
\left(
[\phi(\mathcal{D});\psi(\mathbf{c})]
\right).
\end{equation}

Let $\mathcal{A}=\{\mathbf{c}_1,\ldots,\mathbf{c}_{A}\}$ denote the set of
anchor configurations evaluated during offline meta-training. For an unseen
target dataset $\mathcal{D}_{\mathrm{target}}$, we score each anchor
configuration and select the top-$B$ ranked candidates:
\begin{equation}
\mathcal{C}_{\mathrm{seed}}
=
\operatorname{Top}_{B}
\left(
\mathcal{A};
s_{\mathrm{meta}}(\cdot,\mathcal{D}_{\mathrm{target}})
\right),
\end{equation}

where $\operatorname{Top}_{B}$ returns the $B$ anchor configurations with the
highest predicted scores. The resulting seed set $\mathcal{C}_{\mathrm{seed}}$ provides the starting
configurations for warm-started \mbox{LATS-IR}, transferring promising
configurations from prior retrieval tasks. When
$\mathcal{C}_{\mathrm{seed}}=\varnothing$, the agent instead generates
cold-start configurations; otherwise, the meta-learned seed configurations
initialize the search. Full details of the source benchmarks, anchor
configuration set, meta-feature construction, and meta-learner training and
selection are provided in Appendix~\ref{app:meta_learner}.

\begin{table*}[t]
\centering
\caption{
Characteristics of the target retrieval benchmarks. We report dataset domain, corpus size,
optimization and held-out query counts, and meta-feature distance
$d_{\mathrm{meta}}$. 
}
\label{tab:dataset_stats}

\fontsize{10}{11}\selectfont
\setlength{\tabcolsep}{4pt}
\renewcommand{\arraystretch}{1.05}

\begin{tabular}{@{}l l r r r c@{}}
\toprule
\textbf{Dataset}
& \textbf{Domain}
& \textbf{Corpus Size}
& \textbf{Opt. Queries}
& \textbf{Held-out Queries}
& $\boldsymbol{d_{\mathrm{meta}}}$ \\
\midrule

\texttt{SciFact}
& Scientific fact verification
& 5,183
& 240
& 60
& 0.679 \\

\texttt{FiQA-2018}
& Financial QA
& 57,638
& 519
& 129
& 0.354 \\

\texttt{CQADupStack-STATS}
& Statistics duplicate questions
& 42,269
& 522
& 130
& 0.396 \\

\texttt{CoSQA}
& Natural-language code search
& 6,267
& 400
& 100
& 2.092 \\

\bottomrule
\end{tabular}
\end{table*}

\section{Experimental Setup} \label{sec:experiments}

We evaluate \mbox{IRAH} on four retrieval benchmarks spanning diverse
domains. We compare warm- and cold-started variants of \mbox{IRAH} and Tree-structured Parzen Estimator TPE, together with Grid Search, using a shared retrieval engine and evaluation
protocol.
We optimize all methods using $\mathrm{nDCG}@10$ on an optimization-query split and report performance on a held-out query set. Additional implementation details are in Appendix~\ref{app:exper_setup}.

\subsection{Datasets}
\label{subsec:datasets}

We evaluate \mbox{IRAH} on three diverse, in-distribution natural-language retrieval
benchmarks: \texttt{SciFact} \citep{wadden-et-al-2020-fact},
\texttt{FiQA-2018} \citep{maia2018fiqa}, and
\texttt{CQADupStack-STATS}
\citep{thakur2021beir,hoogeveen2015cqadupstack}.
These benchmarks span scientific fact verification, financial question
answering, and duplicate-question retrieval, respectively.
We additionally evaluate on \texttt{CoSQA}
\citep{huang-etal-2021-cosqa,li-etal-2025-coir}, 
a code-retrieval benchmark
that pairs natural-language queries with source code, 
as an out-of-distribution (OOD) target. This setup allows us to evaluate both performance on targets
similar to the meta-training distribution and transfer under a different retrieval distribution. Table~\ref{tab:dataset_stats} summarizes the benchmark characteristics.

To quantify benchmark-to-meta-training distribution shift, we use the same dataset
representation $\phi(\mathcal{D})$ introduced in
Section~\ref{subsec:meta_learner}. Each feature is normalized using
statistics computed exclusively from the meta-training datasets. Let
$\tilde{\mathbf{x}}_t$ and $\tilde{\mathbf{x}}_m$ denote the normalized
meta-feature vectors of the target dataset and meta-training dataset $m$,
respectively. We define the meta-feature distance as the Euclidean distance
to the nearest meta-training dataset:
\begin{equation}
d_{\mathrm{meta}}
=
\min_{m \in \mathcal{D}_{\mathrm{meta}}}
\left\|
\tilde{\mathbf{x}}_t-\tilde{\mathbf{x}}_m
\right\|_2 .
\end{equation}

The OOD benchmark, \texttt{CoSQA}, is
substantially farther from the meta-training distribution
($d_{\mathrm{meta}}=2.092$) than the three in-distribution targets:
\texttt{SciFact} ($0.679$), \texttt{FiQA-2018} ($0.354$), and
\texttt{CQADupStack-STATS} ($0.396$). Complete meta-feature definitions and
normalization details are provided in Appendix~\ref{app:meta_learner}.

For each dataset, we partition the evaluation queries into an
approximately 80/20 optimization and held-out evaluation split.
All configuration search and selection decisions are made exclusively using
the optimization set $Q_{\mathrm{opt}}$, while the held-out queries are used
only for final evaluation of the selected configuration.




\begin{table*}[t]
\centering
\caption{
Held-out retrieval performance across target benchmarks.
Results are mean $\pm$ standard deviation over three seeds. Grid Search is
reported from a single run. Warm variants use meta-learned initialization,
whereas Cold variants do not. CoSQA is the out-of-distribution (OOD) target.
Average is the unweighted mean held-out $\mathrm{nDCG}@10$ across all
four benchmarks. Best result per dataset is shown in bold.
}
\label{tab:retrieval_results}

\fontsize{9}{10}\selectfont
\setlength{\tabcolsep}{4pt}
\renewcommand{\arraystretch}{1.05}

\begin{tabular}{@{}lccccc@{}}
\toprule
\textbf{Method}
& \textbf{SciFact}
& \textbf{FiQA}
& \shortstack{\textbf{CQADupStack-}\\\textbf{STATS}}
& \shortstack{\textbf{CoSQA}\\\textbf{(OOD)}}
& \shortstack{\\\textbf{Average}} \\
\midrule

Grid Search
& 0.7321
& 0.4407
& 0.3602
& 0.3350
& 0.4670 \\

TPE (Cold)
& $0.7158 \pm .0035$
& $0.4843 \pm .0152$
& $0.3630 \pm .0053$
& $0.3181 \pm .0090$
& 0.4703 \\

TPE (Warm)
& $0.7176 \pm .0113$
& $0.4662 \pm .0198$
& $\mathbf{0.3675 \pm .0036}$
& $\mathbf{0.3420 \pm .0050}$
& 0.4733 \\

IRAH (Cold)
& $0.7267 \pm .0210$
& $0.4835 \pm .0186$
& $0.3542 \pm .0111$
& $0.3088 \pm .0222$
& 0.4683 \\

IRAH (Warm)
& $\mathbf{0.7368 \pm .0075}$
& $\mathbf{0.4959 \pm .0021}$
& $0.3601 \pm .0013$
& $0.3336 \pm .0131$
& $\mathbf{0.4816}$ \\

\bottomrule
\end{tabular}
\end{table*}

\subsection{IRAH Implementation}
\label{subsec:implementation}

We instantiate \mbox{IRAH} as follows. For warm-start initialization, we use
an \texttt{XGBRanker} meta-learner trained on 11 source retrieval benchmarks
using performance measurements from a 36-configuration anchor set. Each
dataset--configuration pair is represented by 13 dataset-level meta-features
and 14 configuration features. At inference, \texttt{XGBRanker} scores the
36 anchor configurations for the target dataset and returns the top $B=3$
configurations for warm-start initialization. Complete meta-training details,
including the source benchmarks, feature definitions, anchor configurations,
and model-selection procedure, are provided in
Appendix~\ref{app:meta_learner}. 

The downstream search space spans four chunking strategies
(\texttt{none}, \texttt{token}, \texttt{sentence}, and
\texttt{recursive\_character}); three dense embedding models
(\texttt{multi-qa-mpnet-base-cos-v1}, \texttt{bge-large-en-v1.5}, and
\texttt{bge-m3}); three cross-encoder rerankers plus a no-reranking option;
and numerical parameters controlling chunk size, retrieval depth $k$, and
hybrid lexical--dense fusion weight $\alpha$. Phase~1 searches over pipeline
structure together with coarse numerical values
$k\in\{10,20,30,50\}$ and
$\alpha\in\{0.00,0.33,0.67,1.00\}$.
Phase~2 fixes the selected categorical architecture and performs localized
refinement of chunk size, $k$, and $\alpha$. Complete search-space
specifications and Phase~2 refinement bounds are provided in
Appendix~\ref{app:search_space}.

The fixed search budget ($N_{\max}=50$) is divided between macro-structural exploration and
micro-level refinement using a phase-allocation fraction $\gamma$.
Warm-started runs on the in-distribution benchmarks use $\gamma=0.24$, allocating
12 of 50 evaluations to Phase~1 and 38 to Phase~2, whereas cold-started runs
use $\gamma=0.60$ (30 and 20 evaluations, respectively). To reduce reliance
on transferred configurations under substantial distribution shift,
warm-started targets with $d_{\mathrm{meta}}>1.5$ use the cold-start
allocation. We treat this threshold as a heuristic experimental decision rule
rather than a calibrated OOD boundary. Thus, warm-started in-distribution search
devotes more of its budget to local refinement, while cold-started and OOD
search retain greater structural exploration.



Node selection follows the adaptive UCT procedure described in
Section~\ref{subsec:lats_reflection}. Reflection and candidate mutation are
performed using \texttt{openai/gpt-oss-120b} \cite{openai2025gptoss}, and generated candidates are
validated against the phase-specific search-space constraints before
evaluation. Additional UCT, prompting, and search-memory details are provided
in Appendices~\ref{app:adaptive_uct} and
\ref{app:reflection_mutation}.

\subsection{Baselines and Comparison Protocol}
\label{subsec:methods}

We benchmark \mbox{IRAH} against two standard configuration-optimization
baselines: a fixed 108-configuration Grid Search
(Appendix Table~\ref{tab:appendix_all_108_configs}) and Tree-structured Parzen
Estimator (TPE) \citep{bergstra2011algorithms}, implemented with Optuna
\citep{akiba2019optuna}. All methods evaluate configurations using the same
retrieval engine, query splits, and retrieval objective.

We evaluate both \mbox{IRAH} and TPE under \emph{cold} and \emph{warm}
initialization. Cold variants begin without meta-learned configuration
seeds, whereas warm variants receive the same top-$B$ configurations
proposed by the offline meta-learner. In particular, the shared warm-start
initialization enables \mbox{LATS-IR} and TPE to be compared while holding
the initialization strategy fixed.


Both \mbox{IRAH} and TPE receive a fixed optimization budget of
$N_{\max}=50$ configuration evaluations. Grid Search deterministically
evaluates its predefined configuration set and is therefore reported as a
single run. The stochastic methods are evaluated using three independent
random seeds, and results are reported as mean $\pm$ standard deviation.


\subsection{Evaluation Metrics}
\label{subsec:metrics}

The optimization objective for all methods is $\mathrm{nDCG}@10$ computed
over $Q_{\mathrm{opt}}$. After optimization, the configuration selected by
each run is evaluated on the corresponding held-out query set, and we report
held-out $\mathrm{nDCG}@10$ as the retrieval performance metric.

We additionally report optimization-loop runtime, measured as the wall-clock time attributed to each method's search and evaluation procedure.
Because structural configuration changes can require distinct
physical retrieval indexes, we also analyze the number of unique index
collections constructed during optimization. For \mbox{IRAH}, we further report
LLM token usage as a measure of agentic optimization overhead.

\section{Results and Analysis}
\label{sec:results}

\begin{table}[t]
\centering
\caption{
Method runtime (hours). Results are mean $\pm$ standard
deviation over three seeds. Grid Search is a single run. Lower is better. Best per dataset is shown in bold.
}
\label{tab:runtime_results}

\fontsize{6.4}{7.1}\selectfont
\setlength{\tabcolsep}{2.2pt}
\renewcommand{\arraystretch}{1.08}

\begin{tabular}{@{}lrrrr@{}}
\toprule
\textbf{Method}
& \textbf{SciFact}
& \textbf{FiQA}
& \shortstack{\textbf{CQADupStack-}\\\textbf{STATS}}
& \shortstack{\textbf{CoSQA}\\\textbf{(OOD)}} \\
\midrule

Grid Search
& 6.75
& 56.92
& 42.83
& 18.02 \\

TPE (Cold)
& $5.73 \pm .20$
& $50.89 \pm 3.19$
& $37.11 \pm 4.14$
& $2.06 \pm .47$ \\

TPE (Warm)
& $5.62 \pm .41$
& $\mathbf{44.71 \pm 4.72}$
& $38.73 \pm 2.65$
& $\mathbf{1.77 \pm .03}$ \\

IRAH (Cold)
& $5.99 \pm 1.15$
& $47.41 \pm 6.72$
& $59.92 \pm 2.75$
& $5.69 \pm 2.93$ \\

IRAH (Warm)
& $\mathbf{3.37 \pm .11}$
& $74.49 \pm 20.06$
& $\mathbf{33.28 \pm 2.53}$
& $2.89 \pm .95$ \\

\bottomrule
\end{tabular}
\end{table}

We first analyze retrieval performance, including out-of-distribution
generalization (Section~\ref{sec:performance}), followed by runtime efficiency
(Section~\ref{sec:runtime}) and qualitative search behavior
(Section~\ref{sec:qual_search}). Additional analyses, including
meta-initialization performance, search trajectories, and runtime and index
diagnostics, are provided in Appendix~\ref{app:extra_results}.

\subsection{Retrieval Performance}
\label{sec:performance}

Table~\ref{tab:retrieval_results} compares held-out retrieval quality across
Grid Search, TPE, and \mbox{IRAH}. Because \mbox{IRAH (Warm)} and TPE (Warm)
receive the same meta-learned initialization, their results provide a direct
comparison of the downstream optimization procedures under a shared warm
start. Across all four benchmarks, \mbox{IRAH (Warm)} achieves the highest
unweighted average held-out $\mathrm{nDCG}@10$ of $0.4816$.

Across the three in-distribution benchmarks, \mbox{IRAH (Warm)} achieves the
highest held-out $\mathrm{nDCG}@10$ on \texttt{SciFact} and
\texttt{FiQA-2018}, improving over TPE (Warm) by $2.7\%$ ($+0.0192$) and
$6.4\%$ ($+0.0297$), respectively. On \texttt{CQADupStack-STATS},
\mbox{IRAH (Warm)} remains competitive, trailing TPE (Warm) by $2.0\%$
($-0.0074$). The weaker result on \texttt{CQADupStack-STATS} coincides with
narrower structural exploration: \mbox{IRAH (Warm)} evaluates
$15.7 \pm 3.2$ unique physical collections, compared with
$20.7 \pm 1.5$ for TPE (Warm), and all three \mbox{IRAH} runs favor similar
embedding and chunking choices
(Appendix~\ref{app:runtime_index}, Table~\ref{tab:best_configurations}).
This pattern is consistent with greater exploitation of the meta-learned
initialization, although these diagnostics do not establish a causal
explanation for the performance gap.

The warm-start setting also yields consistently higher performance than the
cold-start setting for \mbox{IRAH} on all three in-distribution benchmarks.
Held-out $\mathrm{nDCG}@10$ increases by $0.0101$, $0.0124$, and $0.0059$
on \texttt{SciFact}, \texttt{FiQA-2018}, and
\texttt{CQADupStack-STATS}, respectively. Per-evaluation analyses further show that the meta-learned initialization does
not account for the gains of warm-started \mbox{IRAH} alone. \mbox{LATS-IR} improves upon the best warm-start seed on every benchmark, with
the largest gain on the OOD \texttt{CoSQA} target. The source of these gains is
dataset dependent. Local Phase~2 refinement contributes most on
\texttt{FiQA-2018} and \texttt{CQADupStack-STATS}, whereas broader Phase~1
exploration drives the improvement on \texttt{CoSQA}
(Appendix~\ref{app:search_trajectories}).

\paragraph{Out-of-Distribution Generalization}

We further evaluate transfer to \texttt{CoSQA}, our OOD benchmark. Although all four evaluation benchmarks are held out from
meta-learner training, \texttt{CoSQA} lies substantially farther from the
meta-training distribution, with $d_{\mathrm{meta}}=2.092$
(Table~\ref{tab:dataset_stats}). Consistent with this larger distribution
shift, the meta-learner's top-ranked anchor configuration achieves only
$0.2593$ held-out $\mathrm{nDCG}@10$ on \texttt{CoSQA}
(Appendix~\ref{app:meta_initialization}), substantially below its direct
transfer performance on the in-distribution targets.

Despite this weaker transfer signal, meta-learned initialization benefits downstream search. \mbox{IRAH (Warm)} improves from $0.3088$ to
$0.3336$ held-out $\mathrm{nDCG}@10$ over \mbox{IRAH (Cold)}, a gain of
$0.0248$ ($8.0\%$). Although TPE (Warm) achieves the highest score on this
benchmark ($0.3420$), \mbox{IRAH (Warm)} remains competitive and substantially
improves over its cold-start counterpart. This suggests that \mbox{LATS-IR} can recover strong performance from
weaker meta-learned initialization, and that transferred configuration
information can remain useful under distribution shift even when direct
meta-learner recommendations are less reliable.



\subsection{Runtime Efficiency}
\label{sec:runtime}
Method runtime varies across datasets. As shown in
Table~\ref{tab:runtime_results}, \mbox{IRAH (Warm)} is the fastest method
on \texttt{SciFact} and \texttt{CQADupStack-STATS}, while TPE is faster
on \texttt{CoSQA}. On \texttt{SciFact}, \mbox{IRAH (Warm)} reduces
method runtime by approximately 50\% relative to Grid Search.
\texttt{FiQA-2018} incurs substantially greater runtime for
\mbox{IRAH (Warm)}. Appendix~\ref{app:runtime_index} further analyzes
this behavior using physical index usage and pure search/evaluation time,
showing that the increase is not explained by collection count alone.

\subsection{Qualitative Search Behavior}
\label{sec:qual_search}

\begin{figure}[t]
\centering
\includegraphics[width=\linewidth]{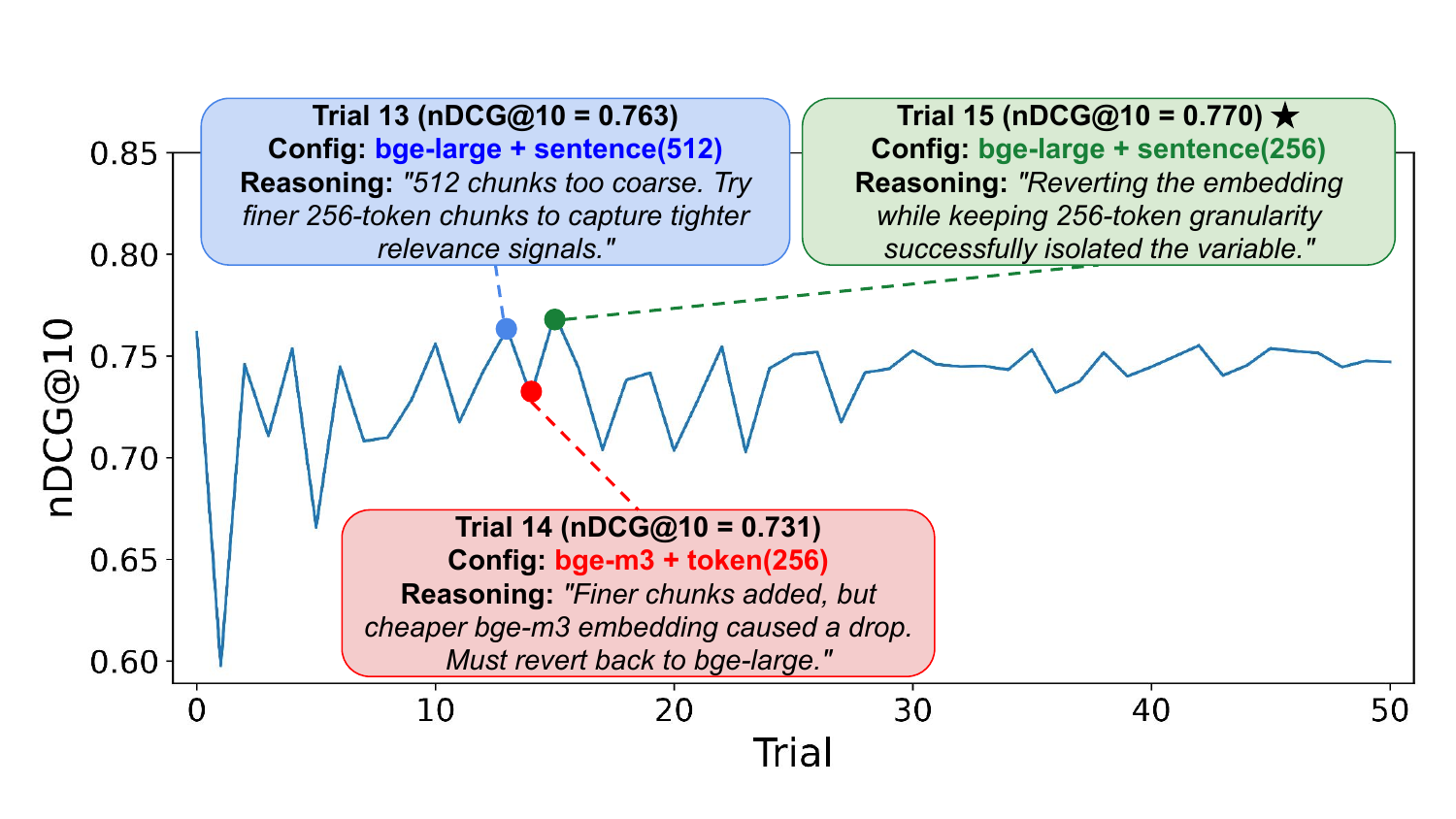}
\caption{
Representative IRAH 
reflection trace on
\texttt{SciFact}. An unsuccessful mutation is followed
by a reflection-guided corrective mutation that reaches a new trajectory peak.
}
\label{fig:irah_self_debugging_scifact}
\end{figure}


Beyond retrieval performance, \mbox{IRAH} retains natural-language
reflections that provide an inspectable record of the search process.
Figure~\ref{fig:irah_self_debugging_scifact} shows a representative
three-trial sequence on \texttt{SciFact}. After a strong configuration at
Trial~13 ($\text{nDCG}@10=0.763$), Trial~14 jointly changes chunk size and
embedding model, reducing performance to $0.731$. The subsequent reflection
identifies the embedding-model change as a likely source of degradation and
motivates a more targeted mutation.
At Trial~15, the search restores \texttt{bge-large} while retaining finer
chunk granularity, increasing nDCG@10 to $0.770$ and establishing a new
trajectory peak. 

Although multiple configuration choices still change across
the sequence, the trace illustrates how stored reflections can redirect search
after an unsuccessful mutation while exposing the rationale behind subsequent
decisions. Such process-level traceability is particularly valuable in
real-world deployments, where inspectable records of system decisions can
support human oversight, debugging, and internal auditing
\citep{raji2020closing}.

\section{Conclusion}

We introduced \mbox{IRAH}, an agentic framework for IR configuration
optimization that combines \mbox{LATS-IR}, a reflection-guided hierarchical
search procedure, with meta-learned warm-start initialization. A key aspect is that \mbox{LATS-IR} organizes search into
macro-structural exploration followed by local numerical refinement, with natural-language reflections to inform subsequent mutations
and provide an inspectable trace.
Across all four benchmarks, warm-started \mbox{IRAH} achieves the highest average held-out $\mathrm{nDCG}@10$ ($0.4816$) and outperforms its cold-start counterpart. These results show that hierarchical agentic search with meta-learned warm starts is a competitive approach to retrieval pipeline optimization under constrained evaluation budgets.

\section*{Limitations}

Our evaluation covers four target benchmarks and one retrieval backend. Although these benchmarks span scientific, financial,
community-question, and code-retrieval domains, broader evaluation across
retrieval systems, larger collections, and additional domains is needed to
establish how consistently \mbox{IRAH}'s search behavior generalizes.

\mbox{LATS-IR} relies on an LLM for reflection and candidate mutation
throughout optimization. Our experiments use only \texttt{gpt-oss-120b} and we do not evaluate sensitivity to the choice of language model. This incurs
inference overhead, consuming roughly 0.6 million tokens per 50-trial run.
Future work could investigate smaller specialized models or distilled
reflection policies while preserving the inspectability of the search
trajectory.



Finally, warm-start effectiveness depends on how well the meta-training
benchmarks represent the target distribution. 
Our OOD experiment shows that meta-learned initialization can remain useful
under distribution shift, though direct-transfer quality is weaker than for the in-distribution targets.
Expanding the
diversity of meta-training datasets and incorporating uncertainty into
warm-start selection are important directions for improving robustness.


\bibliography{custom}

\newpage
\clearpage

\appendix
\section{Additional Experimental Setup}\label{app:exper_setup}

\subsection{Meta-Learner Instantiation}
\label{app:meta_learner}

To construct the warm-start prior, we evaluate three offline meta-learning
models across 11 public IR benchmarks
(Appendix~\ref{app:meta-training-datasets}) using Group $K$-Fold
cross-validation grouped by dataset ID to prevent cross-domain leakage.
Offline training data are generated by evaluating 36 candidate retrieval
configurations on each source dataset
(Appendix~\ref{app:offline-meta-training}).

Each dataset is represented by 13 corpus- and query-level meta-features
(Appendix~\ref{app:meta-feature-rep}), which are concatenated with 14
configuration-specific features to form a 27-dimensional representation for
each dataset--configuration pair. The meta-learner is trained to score candidate configurations according to the target retrieval metric
(i.e., $\text{nDCG}@10$), aligning predicted seed
configurations with the downstream optimization objective.

Among the evaluated models, \texttt{XGBRanker} achieves one of the highest
out-of-fold Spearman-$\rho$ correlations and is selected as the final
meta-learner (Appendix~\ref{app:meta-learner-model}). It is subsequently
retrained using the full 11-dataset meta-training corpus. For an unseen target
dataset at inference time, the system extracts its 13-dimensional meta-feature representation,
scores the 36 candidate anchor configurations, and returns the top-3 ranked
configurations for warm-start initialization.
\subsubsection{Meta-Training Benchmarks}
\label{app:meta-training-datasets}

The meta-training corpus comprises 11 public IR benchmarks spanning diverse
domains and scales. These include \texttt{SciDocs}
\citep{cohan2020specter,thakur2021beir}, \texttt{ArguAna}
\citep{wachsmuth2018retrieval,thakur2021beir}, \texttt{NFCorpus}
\citep{boteva2016nfcorpus,thakur2021beir}, \texttt{Vaswani} \citep{vaswani1970npl},
\texttt{Cranfield} \citep{cleverdon1967cranfield}, and six
\texttt{CQADupStack} subforums
\citep{hoogeveen2015cqadupstack,thakur2021beir}:
\texttt{Physics}, \texttt{GIS}, \texttt{Programmers}, \texttt{Android},
\texttt{Mathematica}, and \texttt{Webmasters}.
\subsubsection{Offline Training Data Collection}
\label{app:offline-meta-training}

To generate ground-truth utility targets, we evaluate each meta-training
benchmark across a targeted 36-configuration subset
($\text{cfg\_01}$--$\text{cfg\_36}$ in
Table~\ref{tab:appendix_all_108_configs}). These configurations were selected
to provide broad coverage of the retrieval search space, spanning six physical
index collections, multiple dense embedding models and document chunking
strategies, and varied numerical parameters ($\alpha$, $k$) and reranker
architectures, while avoiding the computational cost of exhaustive evaluation. The resulting meta-training corpus contains 396 dataset--configuration
pairs from 11 source datasets and 36 anchor configurations.


\subsubsection{Meta-Feature and Configuration Representation}
\label{app:meta-feature-rep}

Each dataset is represented by a 13-dimensional meta-feature vector capturing
corpus, query, and query--corpus characteristics. The features comprise:
(i)~\textbf{Embedding Similarity Statistics}
(\texttt{mean\_similarity}, \texttt{similarity\_std}, and
\texttt{similarity\_moe});
(ii)~\textbf{Query--Corpus Interaction}
(\texttt{query\_vocab\_overlap} and \texttt{avg\_query\_length});
(iii)~\textbf{Document and Length Statistics}
(\texttt{avg\_doc\_length\_words}, \texttt{doc\_len\_std},
\texttt{doc\_len\_cv}, \texttt{char\_per\_word\_ratio}, and
\texttt{query\_doc\_len\_ratio}); and
(iv)~\textbf{Lexical Diversity}
(\texttt{vocab\_entropy}, \texttt{type\_token\_ratio}, and
\texttt{query\_vocab\_entropy}).

Lexical statistics are computed from tokenized query and corpus text.
Query--corpus vocabulary overlap is
\begin{equation}
\frac{|V_q \cap V_c|}{|V_q|},
\end{equation}
where $V_q$ and $V_c$ denote the unique query and corpus vocabularies.
Document-length coefficient of variation and the query--document length ratio
are respectively
\begin{equation}
\frac{\sigma_{\ell_d}}{\mu_{\ell_d}}
\quad\text{and}\quad
\frac{\mu_{\ell_q}}{\mu_{\ell_d}},
\end{equation}
where $\ell_q$ and $\ell_d$ denote query and document lengths.
Type--token ratio is $|V_c|/N_c$, where $N_c$ is the total number of corpus
tokens, and lexical entropy is computed as
\begin{equation}
H(V)=-\sum_{w\in V}p(w)\log p(w).
\end{equation}
Embedding-similarity features summarize query--corpus semantic similarities
using their empirical mean, standard deviation, and margin-of-error estimate.

For each candidate configuration, the dataset representation is concatenated
with 14 configuration-level features: fusion weight $\alpha$, retrieval depth
$k$, chunk size, three embedding-model indicators, four reranker indicators,
and four chunking-strategy indicators. This yields a 27-dimensional feature
vector for each dataset--configuration pair.

At inference, the target dataset's 13-dimensional meta-feature vector is paired
with each of the 36 anchor configurations. The trained meta-learner scores all
36 pairs, and the top $B=3$ configurations by predicted utility are used to
initialize warm-started \mbox{LATS-IR}.

\subsubsection{Cross-Validation and Model Selection}
\label{app:meta-learner-model}

We compare three offline meta-learning models for predicting per-configuration
retrieval utility: (1)~\texttt{XGBRanker} with a \texttt{rank:pairwise} objective, 
(2)~\texttt{Random Forest Regressor}, 
and
(3)~\texttt{Ridge Regression}. We note the meta-learning formulation is model-agnostic. In our
experiments, configuration utility is defined by the downstream optimization
metric (i.e., $\text{nDCG}@10$).

To prevent cross-dataset leakage, hyperparameter tuning uses 5-fold
Group $K$-Fold cross-validation grouped by dataset ID, ensuring that all
configuration evaluations from a dataset remain within the same fold.


Hyperparameters are selected by grid search using dataset-grouped
cross-validation and out-of-fold Spearman rank correlation ($\rho$) between
predicted and observed configuration utilities:
\begin{itemize}
    \item \textbf{\texttt{XGBRanker} (54 combinations):}
    $\texttt{max\_depth}\in\{3,4,5\}$,
    $\texttt{learning\_rate}\in\{0.01,0.05,0.1\}$,
    $\texttt{n\_estimators}\in\{50,100,200\}$, and
    $\texttt{subsample}\in\{0.8,1.0\}$, with
    $\texttt{colsample\_bytree}=0.8$ and a pairwise ranking objective.

    \item \textbf{\texttt{Regression Baselines:}}
    Random Forest (27 combinations) and Ridge
    ($\alpha\in\{0.01,0.1,1.0,10.0,100.0\}$) are tuned using the same
    grouped cross-validation protocol.
\end{itemize}

\begin{table}[h]
\centering
\small
\caption{
Out-of-fold Spearman rank correlation ($\rho$) under dataset-grouped
cross-validation for candidate meta-learners.
}
\label{tab:surrogate_comparison}
\resizebox{\columnwidth}{!}{%
\begin{tabular}{l c l}
\toprule
\textbf{Model} &
\textbf{CV Spearman $\rho$} &
\textbf{Type} \\
\midrule
\texttt{RandomForest}
    & 0.6352 & Pointwise Regressor \\
\texttt{XGBRanker} (Selected)
    & 0.6294 & Pairwise Learning-to-Rank \\
\texttt{Ridge}
    & 0.3772 & Linear Regressor \\
\bottomrule
\end{tabular}%
}
\end{table}

Although \texttt{Random Forest} achieves a marginally higher cross-validation
Spearman correlation ($0.6352$ vs.\ $0.6294$ in Table \ref{tab:surrogate_comparison}), we select
\texttt{XGBRanker} because its pairwise learning-to-rank objective directly
matches the downstream task of ranking configurations for warm-start
selection.
The selected \texttt{XGBRanker} uses
$\texttt{max\_depth}=4$,
$\texttt{learning\_rate}=0.1$,
$\texttt{n\_estimators}=50$,
$\texttt{subsample}=1.0$, and
$\texttt{colsample\_bytree}=0.8$.



To further evaluate cross-dataset generalization, we perform
leave-one-dataset-out evaluation of the selected \texttt{XGBRanker}, training
on 10 source datasets and evaluating its configuration ranking on the held-out
dataset. Table~\ref{tab:meta_learner_loo} reports Spearman correlation together
with ranking-oriented metrics that directly evaluate the quality of the
top-ranked warm-start candidates. Top-1 regret measures the utility gap between the highest-ranked predicted
configuration and the oracle-best configuration; Regret@3 uses the best
configuration among the top three predictions. Hit Rate@3 indicates whether
the oracle-best configuration appears in the predicted top three, while
Recall@3 measures overlap between the predicted and oracle top-three sets.

\begin{table*}[h]
\centering
\caption{
Leave-one-dataset-out evaluation of \texttt{XGBRanker} across the 11
meta-training datasets. Metrics evaluate predicted configuration rankings;
\textbf{Config. nDCG@3} denotes configuration-ranking nDCG rather than
retrieval nDCG.
}
\label{tab:meta_learner_loo}
\resizebox{\linewidth}{!}{%
\begin{tabular}{l c c c c c c}
\toprule
\textbf{Target Dataset} &
\textbf{Spearman $\rho$} &
\textbf{Top-1 Regret} &
\textbf{Regret@3} &
\textbf{Hit Rate@3} &
\textbf{Recall@3} &
\textbf{Config. nDCG@3} \\
\midrule
\texttt{arguana}
    & 0.4432 & 0.0782 & 0.0000 & 1.0000 & 0.3333 & 0.8684 \\
\texttt{cqadupstack/android}
    & 0.9197 & 0.0000 & 0.0000 & 1.0000 & 0.6667 & 0.9977 \\
\texttt{cqadupstack/gis}
    & 0.9794 & 0.0000 & 0.0000 & 1.0000 & 0.6667 & 0.9968 \\
\texttt{cqadupstack/mathematica}
    & 0.9369 & 0.0000 & 0.0000 & 1.0000 & 1.0000 & 1.0000 \\
\texttt{cqadupstack/physics}
    & 0.9799 & 0.0000 & 0.0000 & 1.0000 & 1.0000 & 1.0000 \\
\texttt{cqadupstack/programmers}
    & 0.9669 & 0.0000 & 0.0000 & 1.0000 & 0.6667 & 0.9989 \\
\texttt{cqadupstack/webmasters}
    & 0.8396 & 0.0000 & 0.0000 & 1.0000 & 0.6667 & 0.9908 \\
\texttt{nfcorpus}
    & 0.8108 & 0.0000 & 0.0000 & 1.0000 & 0.6667 & 0.9888 \\
\texttt{scidocs}
    & 0.8583 & 0.0000 & 0.0000 & 1.0000 & 0.3333 & 0.9451 \\
\texttt{cranfield}
    & 0.6166 & 0.0243 & 0.0239 & 0.0000 & 0.0000 & 0.9391 \\
\texttt{vaswani}
    & 0.2462 & 0.0909 & 0.0536 & 0.0000 & 0.0000 & 0.8616 \\
\midrule
\textbf{Mean}
    & \textbf{0.7816}
    & \textbf{0.0176}
    & \textbf{0.0070}
    & \textbf{0.8182}
    & \textbf{0.5455}
    & \textbf{0.9625} \\
\bottomrule
\end{tabular}%
}
\end{table*}

Leave-one-dataset-out evaluation yields a mean Spearman correlation of
$0.7816$ and configuration-ranking $\text{nDCG}@3$ of $0.9625$.
The low mean Regret@3 ($0.0070$) further indicates that high-utility
configurations are generally retained near the top of the predicted ranking,
which is particularly relevant because \mbox{IRAH} uses the top-3 ranked
configurations for warm-start initialization.

\begin{table}[h]
\centering
\small
\caption{
Top-12 feature importances for the trained \texttt{XGBRanker}, using the
default XGBoost sklearn \texttt{feature\_importances} values
(normalized gain for tree-based boosters).
}
\label{tab:feature_importance}
\resizebox{\columnwidth}{!}{%
\begin{tabular}{l c}
\toprule
\textbf{Feature} & \textbf{Normalized Gain Importance} \\
\midrule
\texttt{alpha} ($\alpha$ fusion weight) & 0.0988 \\
\texttt{emb\_bge-large-en-v1.5} & 0.0666 \\
\texttt{reranker\_none} & 0.0607 \\
\texttt{emb\_bge-m3} & 0.0604 \\
\texttt{query\_doc\_len\_ratio} & 0.0593 \\
\texttt{type\_token\_ratio} & 0.0479 \\
\texttt{query\_vocab\_overlap} & 0.0464 \\
\texttt{doc\_len\_cv} & 0.0427 \\
\texttt{mean\_similarity} & 0.0404 \\
\texttt{top\_k} & 0.0402 \\
\texttt{similarity\_std} & 0.0379 \\
\texttt{query\_vocab\_entropy} & 0.0372 \\
\bottomrule
\end{tabular}
}
\end{table}

\begin{table}[h]
\centering
\small
\setlength{\tabcolsep}{4pt}
\caption{
Aggregate feature importance by category for the
\texttt{XGBRanker} meta-learner.
}
\label{tab:category_importance_unbiased}
\resizebox{\columnwidth}{!}{%
\begin{tabular}{lccc}
\toprule
\textbf{Feature Group} &
\textbf{\#} &
\textbf{Cumulative Importance} &
\textbf{Mean Importance} \\
\midrule
Continuous Parameters ($\alpha$, $k$, chunk size)
    & 3  & 0.1636 (16.4\%) & 0.0545 \\
Embedding Model
    & 3  & 0.1572 (15.7\%) & 0.0524 \\
Dataset Structural / Lexical Features
    & 11 & 0.4359 (43.6\%) & 0.0396 \\
Reranker Model
    & 4  & 0.1323 (13.2\%) & 0.0331 \\
Similarity / Diagnostics
    & 2  & 0.0530 (5.3\%)  & 0.0265 \\
Chunking Strategy
    & 4  & 0.0579 (5.8\%)  & 0.0145 \\
\midrule
\textbf{Total / Average}
    & \textbf{27}
    & \textbf{1.0000 (100.0\%)}
    & \textbf{0.0370} \\
\bottomrule
\end{tabular}%
}
\end{table}

Feature-importance analysis (Tables~\ref{tab:feature_importance}--\ref{tab:category_importance_unbiased}) shows that the meta-learner relies on both dataset characteristics and
configuration choices when ranking warm-start candidates. Dataset-level
features account for $48.9\%$ of cumulative normalized gain importance, while
continuous configuration parameters, embedding choice, reranker choice, and
chunking strategy account for the remainder. At the individual-feature level,
the fusion weight $\alpha$ has the highest importance, although importance is
distributed across multiple dataset and configuration features rather than
being dominated by a single factor. 


\subsection{LATS-IR Instantiation}
\label{app:lats_details}

LATS \citep{zhou2023language} applies tree search and reflection to
language-agent task-solving trajectories. In contrast, \mbox{LATS-IR}
operates over complete IR pipeline configurations, with actions corresponding
to structural or numerical configuration mutations. The main formulation is
described in Section~\ref{subsec:lats_reflection}; here we provide the
phase-specific search space and implementation details used in our experiments.

\subsubsection{Search Space and Phase-Specific Refinement}
\label{app:search_space}

\begin{table}[h]
\centering
\small
\setlength{\tabcolsep}{6pt}
\caption{Base pipeline search space used for macro-structural search.
Phase~2 performs localized numerical refinement beyond the discrete values
listed here, as described below.}
\label{tab:appendix_global_search_space}
\resizebox{\columnwidth}{!}{%
\begin{tabular}{@{}lp{0.73\columnwidth}@{}}
\toprule
\textbf{Pipeline Parameter} & \textbf{Option Space / Parameter Bounds} \\
\midrule

\textbf{Chunking Strategy}
& \texttt{none} (size: 0, overlap: 0) \\
& \texttt{token} (256/25, 512/50, 512/100) \\
& \texttt{sentence} (256/25, 512/50, 512/100) \\
& \texttt{recursive\_character} (1500/150, 3000/300, 3000/600) \\

\midrule

\textbf{Dense Embedding Model}
& \texttt{multi-qa-mpnet-base-cos-v1} \citep{song2020mpnet} \\
& \texttt{BAAI/bge-large-en-v1.5} \citep{xiao2023c} \\
& \texttt{BAAI/bge-m3} \citep{chen2024bge} \\

\midrule

\textbf{Cross-Encoder Reranker}
& \texttt{ms-marco-MiniLM-L-6-v2} \citep{reimers2019sentence} \\
& \texttt{BAAI/bge-reranker-large} \citep{xiao2023c} \\
& \texttt{BAAI/bge-reranker-v2-m3} \citep{chen2024bge} \\
& \texttt{none} (no reranking) \\

\midrule

\textbf{Retrieval Depth ($k$)}
& Phase~1 grid: $k \in \{10,20,30,50\}$ \\

\midrule

\textbf{Fusion Weight ($\alpha$)}
& Phase~1 grid: $\alpha \in \{0.00,0.33,0.67,1.00\}$ \\

\bottomrule
\end{tabular}%
}
\end{table}

The joint pipeline search space combines categorical
macro-structural choices $\mathcal{M}$ with numerical parameters $\Theta$.
Table~\ref{tab:appendix_global_search_space} specifies the discrete base
space used during structural search. Phase~2 subsequently permits localized
off-grid refinement of selected numerical parameters.

In all experiments, \mbox{LATS-IR} evaluates a fixed budget of 50 trials.
Tree depth is capped at 18 as a safety guardrail, while node selection within
the tree is governed by UCT. The two search phases differ in the type and
range of configuration mutations permitted:

\begin{itemize}

    \item \textbf{Phase~1 (Macro-Structural Search):}
    The optimizer searches over categorical pipeline structure, including
    chunking strategy, embedding model, and reranker. Numerical parameters
    are restricted to the coarse grids
    $k \in \{10,20,30,50\}$ and
    $\alpha \in \{0.00,0.33,0.67,1.00\}$.
    All categorical choices in
    Table~\ref{tab:appendix_global_search_space} are eligible for mutation.

    \item \textbf{Phase~2 (Micro-Numerical Refinement):}
    The categorical identity of the best-performing Phase~1 architecture is
    held fixed, while \mbox{LATS-IR} performs localized numerical refinement
    around its Phase~1 parameters.
    \textbf{Chunk size} may vary within a $\pm128$ token/character window
    around the Phase~1 value and is not restricted to the discrete values in
    Table~\ref{tab:appendix_global_search_space}. Chunk overlap is not
    independently optimized; it is obtained from the canonical table when
    available or otherwise set using
    $\lfloor \text{chunk\_size}/5 \rfloor$.
    \textbf{Retrieval depth $k$} is adjusted within a $\pm10$ window around
    the Phase~1 value, bounded to $[10,50]$.
    \textbf{Fusion weight $\alpha$} is refined using an initial
    $\pm0.15$ local window that narrows to $\pm0.05$ after three refinement
    cycles, with step size $\Delta=0.05$.

\end{itemize}

\subsubsection{Adaptive UCT Exploration Weight}
\label{app:adaptive_uct}

Rather than using a fixed exploration weight $w$ in
Equation~\ref{eq:uct}, \mbox{LATS-IR} recomputes $w_t$ before each node
selection using a bias-corrected Exponentially Weighted Moving Average (EWMA)
of observed trial rewards, with decay factor $\beta=0.9$.
The resulting weight controls the UCT exploration--exploitation trade-off
within each search phase and is distinct from the structural-versus-numerical
distinction between Phase~1 and Phase~2.

Let $\bar{R}_{\mathrm{EWMA},t}$ denote the bias-corrected EWMA reward.
We use phase-specific schedules:

\begin{itemize}

    \item \textbf{Phase~1 (Macro-Structural Search):}
    \[
    \begin{aligned}
    w_t &= 2.0 - 0.3\,\bar{R}_{\mathrm{EWMA},t},\\
    w_t &\in [1.7,2.0].
    \end{aligned}
    \]
    The comparatively large exploration weight encourages broad traversal of
    alternative structural configurations, while gradually placing greater
    emphasis on higher-valued branches as observed rewards increase.

    \item \textbf{Phase~2 (Micro-Numerical Refinement):}
    \[
    \begin{aligned}
    w_t &= 0.1 + 0.3\left(1-\bar{R}_{\mathrm{EWMA},t}\right),\\
    w_t &\in [0.1,0.4].
    \end{aligned}
    \]
    The lower exploration range emphasizes local refinement around the
    selected Phase~1 architecture. When recent reward estimates are lower,
    $w_t$ increases toward $0.4$, allocating relatively more UCT exploration
    among candidate refinement paths.

\end{itemize}

\subsubsection{Reflection and Mutation Implementation}
\label{app:reflection_mutation}

At each evaluated node, \mbox{LATS-IR} uses an LLM-based reflection and
mutation procedure to inform subsequent candidate generation. After evaluating
a configuration, the reflection prompt provides the model with the evaluated
configuration, its observed retrieval scores, the branch trajectory leading to
the node, and the rationale associated with the proposed configuration. For
non-root nodes, the trajectory includes the parent configuration and reward
together with the observed parent-relative outcome; root nodes are treated as
initial expansions.

The reflection prompt follows the template:
\begin{quote}\small
You are evaluating an IR retrieval configuration for a search optimization
experiment. \\
\textbf{Config:} \{embedding\}, \{chunker\}, $\alpha=\{\alpha\}$,
$k=\{k\}$, \{reranker\}. \\
\textbf{Scores:} \{metric values\}. \\
\textbf{Search branch that led here:} \{branch trajectory and prior
critiques\}. \\
\textbf{Rationale:} \{candidate-generation rationale\}. \\
Write exactly two sentences: (1) summarize what the scores suggest about the
configuration, and (2) suggest what the next expansion should try or avoid.
\end{quote}

The resulting reflection is stored with the search node and can be reused as
context during later candidate generation. Mutation prompts condition on the
UCT-selected branch, previous evaluations, the active search phase, and the
valid choices or bounds for that phase. The shared context includes:
\begin{quote}\small
\textbf{Branch trajectory:} \{selected root-to-leaf path with reflections\}. \\
\textbf{Trial history:} \{previously evaluated configurations and scores\}. \\
\textbf{Observed patterns:} \{summary of prior outcomes\}. \\
\textbf{Objective:} \{optimization objective\}. \\
\textbf{Target score:} \{target reward\}.
\end{quote}

During Phase~1, the prompt permits both structural mutations and coarse
numerical changes:
\begin{quote}\small
\textbf{Phase 1: Structural and Coarse Numerical Exploration.}
Explore combinations of chunking strategy, embedding model, and reranker,
together with the allowed coarse chunk-size, $\alpha$, and $k$ values.
Use only the valid choices provided for this phase, and do not intentionally
repeat an exact configuration from the trial history.
\end{quote}

During Phase~2, the selected categorical structure is held fixed and the
prompt restricts proposals to local numerical refinement:
\begin{quote}\small
\textbf{Phase 2: Local Numerical Refinement.}
Keep the selected chunking strategy, embedding model, and reranker fixed.
Refine $\alpha$ within the active local window, vary $k$ near the selected
value, and adjust chunk size within the allowed range. Avoid previously tried
Phase~2 configurations.
\end{quote}

Candidate configurations are emitted through a structured tool-call interface
rather than free-form text:
\begin{quote}\small
\texttt{propose\_candidate\_eve(embedding\_model, reranker\_model, alpha,
top\_k, chunk\_strategy, chunk\_size, chunk\_overlap, reasoning)}
\end{quote}
After proposing the requested candidates, the model issues a finalization call.
Each proposed configuration is validated against the enumerated model choices,
chunking recipes, and numerical bounds before retrieval evaluation. Invalid
tool arguments are rejected and may be corrected within the bounded tool-call
budget.

\paragraph{Implementation constants.}
Each expansion requests $K=3$ candidate configurations. The proposal agent is
allowed up to six tool-call iterations, permitting invalid tool arguments to
be corrected before finalization. Invalid configurations are not passed to
retrieval evaluation. Stochastic optimization runs use a maximum budget of
50 evaluated configurations and three random seeds.

\paragraph{Duplicate handling.}
The candidate-generation prompt includes the prior trial history and explicitly
instructs the model not to repeat previously evaluated configurations.
Duplicate rejection is not additionally enforced as a hard constraint by the
proposal tool.

All reflection and mutation calls use
\texttt{openai/gpt-oss-120b} \citep{openai2025gptoss}. Reflection and proposal generation uses temperature $0$.

\subsection{Optimization Baselines}
\label{app:baselines}

We benchmark \mbox{IRAH} against two standard optimization baselines, a fixed non-exhaustive Grid Search and Tree-structured Parzen Estimator (TPE) \citep{bergstra2011algorithms}
optimization implemented with Optuna \citep{akiba2019optuna}.

The Grid Search evaluates 108 predefined pipeline configurations spanning
18 physical index collections. Each collection is determined by a specific
dense embedding model and chunking configuration, while reranker choice,
retrieval depth $k$, and fusion weight $\alpha$ vary across configurations.
The complete specification of all 108 configurations is provided in
Table~\ref{tab:appendix_all_108_configs}.

TPE is evaluated under the same 50-trial budget as \mbox{IRAH} and searches
over the same pipeline components, including chunking strategy, embedding
model, reranker, retrieval depth $k$, and fusion weight $\alpha$.
Categorical choices follow those listed in
Table~\ref{tab:appendix_global_search_space}, while numerical parameters are
sampled over their allowed ranges:
$k \in [10,50]$ and
$\alpha \in [0.0,1.0]$.
Cold-started TPE begins without meta-learned initialization, whereas
TPE (Warm) receives the same meta-learned seed initialization used by
\mbox{IRAH (Warm)}.
\subsection{Retrieval Engine and Evaluation Protocol}
\label{app:iris}

All candidate retrieval configurations are evaluated using the same managed
enterprise retrieval engine. Retrieval depth $k$ denotes the number of retrieved
passage chunks rather than aggregated documents. The base search grid uses
$k \in \{10,20,30,50\}$, while optimization methods may propose additional
values within their method-specific numerical bounds
(Section~\ref{app:search_space}).

Hybrid lexical-semantic retrieval is controlled by a fusion weight
$\alpha \in [0,1]$, where $\alpha=0$ corresponds to pure lexical BM25
retrieval and $\alpha=1$ corresponds to pure dense vector retrieval;
intermediate values combine the two retrieval signals.


\section{Additional Results}
\label{app:extra_results}

\begin{figure*}[t]
\centering
\includegraphics[width=\linewidth]{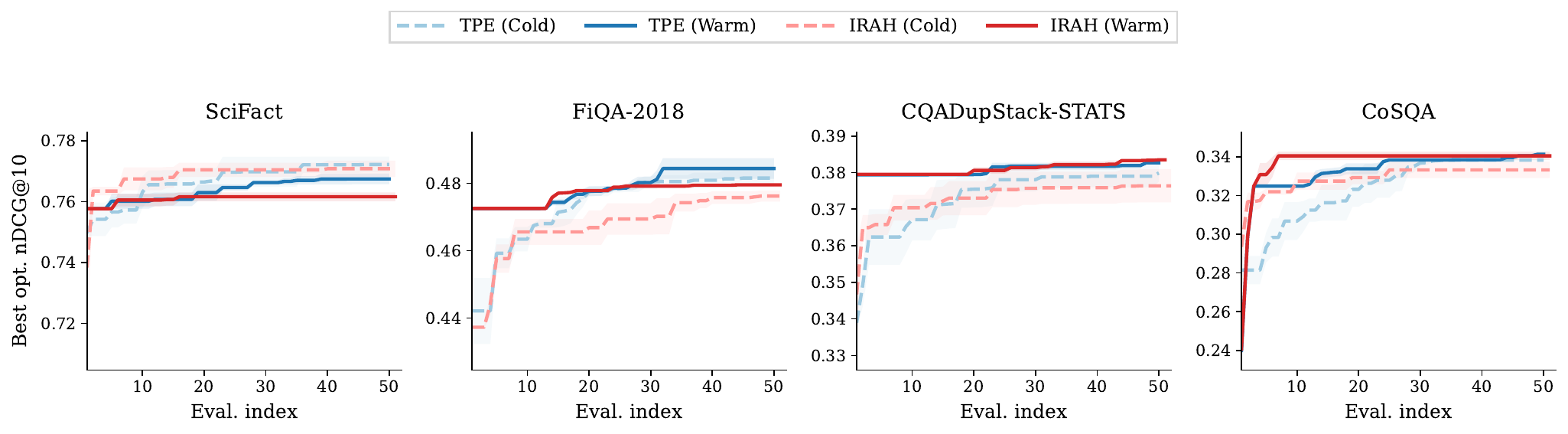}
\caption{
Best-so-far optimization-set $\mathrm{nDCG}@10$ over the 50-evaluation
budget, averaged across three seeds; shading denotes $\pm$ one standard
deviation. Warm variants use the same meta-learned initialization. Held-out
queries are used only for final evaluation in
Table~\ref{tab:retrieval_results}.
}
\label{fig:optimization_trajectories}
\end{figure*}

\subsection{Search Dynamics and Phase Contributions}
\label{app:search_trajectories}
\subsubsection{Optimization Trajectories}
Figure~\ref{fig:optimization_trajectories} complements the final held-out
results in Table~\ref{tab:retrieval_results} by showing how each stochastic
optimizer evolves over the full 50-configuration evaluation budget. Each curve
reports the best optimization-set $\mathrm{nDCG}@10$ observed up to a given
evaluation, averaged across three seeds, with shading indicating one standard
deviation. Thus, the trajectories characterize both the quality of the
initialization and the additional improvement obtained through downstream
search, rather than only the final selected configuration.

The trajectories show that warm-start initialization often places the search
in a strong region early in the optimization process. This effect is
particularly visible on \texttt{FiQA-2018} and
\texttt{CQADupStack-STATS}, where the warm variants begin substantially above
their cold-start counterparts. Subsequent search yields additional gains, but
the magnitude and timing of improvement vary across datasets. On
\texttt{CoSQA}, the out-of-distribution target, \mbox{IRAH (Warm)} rapidly
improves during the first several evaluations despite the weaker direct
transfer of the meta-learned initialization reported in
Appendix~\ref{app:meta_initialization}. This illustrates how downstream search
can compensate for a less reliable warm start under distribution shift.

The figure also shows that optimization-set performance does not directly
determine the final held-out ranking. On \texttt{SciFact} and
\texttt{FiQA-2018}, TPE (Warm) reaches a slightly higher best-so-far
optimization score, whereas \mbox{IRAH (Warm)} achieves the strongest held-out
performance in Table~\ref{tab:retrieval_results}. Conversely, on
\texttt{CQADupStack-STATS}, \mbox{IRAH (Warm)} reaches a comparable or slightly
higher optimization-set incumbent while TPE (Warm) performs better on the
held-out queries. On \texttt{CoSQA}, the two warm methods converge to very
similar optimization-set performance, while TPE (Warm) retains a small
held-out advantage.

These trajectories provide a process-level view of the
optimization behavior that is not visible from final held-out scores alone.
They show how meta-learned initialization and downstream search interact under
a fixed evaluation budget, while also demonstrating that the relative quality
of the configurations selected by the optimizers cannot be inferred solely
from their best observed score on $Q_{\mathrm{opt}}$.

\subsubsection{Improvement Beyond Warm-Start Initialization}
\label{app:warm_search_gain}

\begin{table*}[t]
\centering
\caption{
Optimization-set improvement of warm-started \mbox{IRAH} beyond the
meta-learned initialization. ``Best seed'' is the strongest of the three
meta-learned seed configurations, and ``Final best'' is the strongest
configuration identified over the complete search. Values are mean $\pm$
standard deviation over three runs; gain is computed within each run before
aggregation.
}
\label{tab:warm_search_gain}
\small
\setlength{\tabcolsep}{4pt}
\begin{tabular}{lccc}
\toprule
\textbf{Dataset} &
\textbf{Best seed} &
\textbf{Final best} &
\textbf{Search gain} \\
\midrule
SciFact
& $0.7577\pm.0000$
& $0.7616\pm.0025$
& $+0.0039\pm.0025$ \\

FiQA-2018
& $0.4725\pm.0000$
& $0.4795\pm.0003$
& $+0.0070\pm.0003$ \\

CQADupStack-STATS
& $0.3795\pm.0003$
& $0.3835\pm.0002$
& $+0.0040\pm.0003$ \\

CoSQA
& $0.3249\pm.0000$
& $0.3405\pm.0038$
& $+0.0156\pm.0038$ \\
\bottomrule
\end{tabular}
\end{table*}

Table~\ref{tab:warm_search_gain} quantifies the improvement obtained by
\mbox{LATS-IR} beyond the configurations supplied by the offline meta-learner.
Warm-started \mbox{IRAH} improves over its strongest initialization on all four
benchmarks, with optimization-set gains ranging from $0.0039$ on
\texttt{SciFact} to $0.0156$ on \texttt{CoSQA}. Thus, the performance of
warm-started \mbox{IRAH} is improved by the downstream hierarchical search which continues to identify stronger configurations after initialization.

The largest improvement occurs on the OOD \texttt{CoSQA} benchmark, where
the best meta-learned seed achieves $0.3249$ $\mathrm{nDCG}@10$ and downstream
search increases the incumbent to $0.3405$. This is consistent with the
trajectory analysis above. Although direct transfer is weaker under the larger
distribution shift, subsequent search can substantially improve upon the
transferred initialization.

\begin{table*}[t]
\centering
\caption{
Optimization-set improvement from Phase~2 local numerical refinement.
``Best after P1'' is the highest $\mathrm{nDCG}@10$ observed by the end of
Phase~1, and ``Final best'' is the best value observed over the complete search.
Values are mean $\pm$ standard deviation over three seeds. Phase~2 gain is
computed within each run before aggregation.
}
\label{tab:phase2_gain}
\small
\setlength{\tabcolsep}{3.5pt}
\begin{tabular}{llccc}
\toprule
\textbf{Dataset} & \textbf{IRAH} &
\textbf{Best after P1} &
\textbf{Final best} &
\textbf{P2 gain} \\
\midrule
SciFact
& Cold & $0.7705\pm.0041$ & $0.7708\pm.0046$ & $+0.0004\pm.0006$ \\
& Warm & $0.7605\pm.0035$ & $0.7616\pm.0025$ & $+0.0011\pm.0014$ \\

FiQA-2018
& Cold & $0.4694\pm.0081$ & $0.4762\pm.0028$ & $+0.0068\pm.0054$ \\
& Warm & $0.4725\pm.0000$ & $0.4795\pm.0003$ & $+0.0070\pm.0003$ \\

CQADupStack-STATS
& Cold & $0.3757\pm.0078$ & $0.3764\pm.0078$ & $+0.0007\pm.0008$ \\
& Warm & $0.3795\pm.0003$ & $0.3835\pm.0002$ & $+0.0040\pm.0003$ \\

CoSQA
& Cold & $0.3333\pm.0074$ & $0.3333\pm.0074$ & $+0.0000\pm.0000$ \\
& Warm & $0.3405\pm.0038$ & $0.3405\pm.0038$ & $+0.0000\pm.0000$ \\
\bottomrule
\end{tabular}
\end{table*}

\subsubsection{Phase~2 Refinement Gains}
Table~\ref{tab:phase2_gain} quantifies the additional optimization-set
improvement observed after Phase~1 structural and coarse numerical search.
Phase~2 refinement provides its largest and most consistent gains on
\texttt{FiQA-2018}, improving the Phase~1 incumbent by $0.0068$ and $0.0070$
$\mathrm{nDCG}@10$ for cold- and warm-started \mbox{IRAH}, respectively.
Warm-started \mbox{IRAH} also gains $0.0040$ on
\texttt{CQADupStack-STATS}. Refinement yields smaller improvements on
\texttt{SciFact} and no additional improvement on \texttt{CoSQA}, where the
eventual incumbent is already identified during Phase~1.

These results indicate that the contribution of local numerical refinement is
dataset dependent. On some benchmarks, Phase~1 already identifies the eventual
incumbent, whereas on others Phase~2 yields measurable additional gains without
revisiting the selected categorical structure. Because warm- and cold-started
runs use different phase allocations, these results characterize the observed
search process rather than a controlled phase ablation; different allocations
or larger phase-specific budgets could change the relative contribution of
each stage.


\subsection{Meta-Learned Initialization Quality}
\label{app:meta_initialization}

\begin{table}[h]
\centering
\caption{
Meta-feature distance ($d_{\mathrm{meta}}$), nearest meta-training
benchmark, and held-out performance of the top-ranked
\texttt{XGBRanker} configuration. $\Delta$ denotes the difference relative
to the held-out performance of the configuration selected by Grid Search on
the optimization split.
}
\label{tab:meta_learner_analysis}

\fontsize{7}{8}\selectfont
\setlength{\tabcolsep}{3.2pt}

\begin{tabular}{@{}l c l c c@{}}
\toprule
\textbf{Dataset}
& $d_{\mathrm{meta}}$
& \textbf{Nearest Source}
& \textbf{Seed nDCG@10}
& $\boldsymbol{\Delta}$ \\
\midrule

\texttt{SciFact}
& 0.679
& \texttt{NFCorpus}
& 0.7479
& $+0.0158$ \\

\texttt{FiQA-2018}
& 0.354
& \texttt{Webmasters}
& 0.4883
& $+0.0476$ \\

\texttt{CQADupStack-STATS}
& 0.396
& \texttt{Mathematica}
& 0.3610
& $+0.0008$ \\

\midrule
\texttt{CoSQA}
& 2.092
& \texttt{SciDocs}
& 0.2593
& $-0.0757$ \\

\bottomrule
\end{tabular}
\end{table}

Table~\ref{tab:meta_learner_analysis} characterizes the quality of
meta-learned initialization for the three in-distribution targets and the OOD
\texttt{CoSQA} benchmark. The nearest meta-training dataset under the
13-dimensional representation
(Appendix~\ref{app:meta-feature-rep}) is reported to contextualize each
target's meta-feature distance; the meta-learner itself is trained jointly
across all 11 source datasets.

For the three in-distribution targets, the top-ranked \texttt{XGBRanker}
configuration provides strong held-out retrieval performance and exceeds the
held-out performance of the Grid-selected configuration by $0.0158$,
$0.0476$, and $0.0008$ $\mathrm{nDCG}@10$, respectively. In contrast,
direct transfer is substantially weaker on the OOD \texttt{CoSQA} target. The top-ranked seed achieves $0.2593$ and trails the Grid-selected
configuration by $0.0757$. This contrast is consistent with the substantially
larger meta-feature distance observed for \texttt{CoSQA} and illustrates a
limitation of direct meta-learned transfer under distribution shift.

Although the \texttt{XGBRanker} seed is selected from the 36 anchor
configurations contained within the larger 108-configuration Grid Search
space, the two approaches use different selection criteria. Grid Search
selects the configuration with the highest performance on the optimization
split, whereas the meta-learner ranks anchor configurations using target
dataset characteristics and relationships learned across source benchmarks.
The top-ranked meta-learned seed achieves higher held-out performance than
the Grid-selected configuration on all three in-distribution targets, suggesting
that the learned prior can identify initialization candidates that generalize
well to held-out queries. The \texttt{CoSQA} result, however, shows that this
direct transfer can degrade under larger distribution shifts, motivating
continued downstream exploration by \mbox{LATS-IR}.

\begin{table*}[t]
\centering
\small
\caption{
Selected configurations for each optimization method, together with their
optimization-set and held-out $\mathrm{nDCG}@10$ and pipeline parameters.
For stochastic methods, the configuration is taken from the run with the
highest optimization-set score.
}
\label{tab:best_configurations}

\resizebox{\textwidth}{!}{%
\begin{tabular}{l l c c l l c l c c}
\toprule
\textbf{Dataset}
& \textbf{Method}
& \shortstack{\textbf{Opt.}\\$\mathbf{nDCG@10}$}
& \shortstack{\textbf{Held-out}\\$\mathbf{nDCG@10}$}
& \textbf{Embedding}
& \textbf{Chunker}
& \textbf{Chunk Size}
& \textbf{Reranker}
& $\alpha$
& $k$ \\
\midrule

\multirow{5}{*}{\texttt{SciFact}}
 & Grid Search
 & 0.7629 & 0.7321
 & \texttt{bge-large-en-v1.5}
 & \texttt{sentence}
 & 512
 & \texttt{bge-reranker-large}
 & 0.33 & 20 \\

 & TPE (Cold)
 & 0.7759 & 0.7160
 & \texttt{bge-large-en-v1.5}
 & \texttt{sentence}
 & 256
 & \texttt{bge-reranker-large}
 & 0.70 & 16 \\

 & TPE (Warm)
 & 0.7710 & 0.7182
 & \texttt{bge-large-en-v1.5}
 & \texttt{token}
 & 512
 & \texttt{bge-reranker-large}
 & 0.65 & 10 \\

 & IRAH (Cold)
 & 0.7757 & 0.7122
 & \texttt{bge-large-en-v1.5}
 & \texttt{sentence}
 & 256
 & \texttt{bge-reranker-large}
 & 0.62 & 20 \\

 & IRAH (Warm)
 & 0.7645 & 0.7281
 & \texttt{bge-large-en-v1.5}
 & \texttt{sentence}
 & 512
 & \texttt{bge-reranker-large}
 & 0.80 & 30 \\
\midrule

\multirow{5}{*}{\texttt{FiQA-2018}}
 & Grid Search
 & 0.4784 & 0.4407
 & \texttt{multi-qa-mpnet-base-cos-v1}
 & \texttt{token}
 & 256
 & \texttt{bge-reranker-v2-m3}
 & 1.00 & 50 \\

 & TPE (Cold)
 & 0.4863 & 0.4667
 & \texttt{multi-qa-mpnet-base-cos-v1}
 & \texttt{recursive\_character}
 & 3000
 & \texttt{bge-reranker-v2-m3}
 & 1.00 & 14 \\

 & TPE (Warm)
 & 0.4883 & 0.4523
 & \texttt{multi-qa-mpnet-base-cos-v1}
 & \texttt{sentence}
 & 512
 & \texttt{bge-reranker-v2-m3}
 & 0.90 & 16 \\

 & IRAH (Cold)
 & 0.4793 & 0.4980
 & \texttt{multi-qa-mpnet-base-cos-v1}
 & \texttt{none}
 & 0
 & \texttt{none}
 & 0.75 & 30 \\

 & IRAH (Warm)
 & 0.4797 & 0.4938
 & \texttt{multi-qa-mpnet-base-cos-v1}
 & \texttt{recursive\_character}
 & 2760
 & \texttt{none}
 & 0.79 & 50 \\
\midrule

\multirow{5}{*}{\texttt{CQADupStack-STATS}}
 & Grid Search
 & 0.3803 & 0.3602
 & \texttt{bge-large-en-v1.5}
 & \texttt{none}
 & 0
 & \texttt{none}
 & 0.67 & 30 \\

 & TPE (Cold)
 & 0.3845 & 0.3655
 & \texttt{multi-qa-mpnet-base-cos-v1}
 & \texttt{sentence}
 & 512
 & \texttt{ms-marco-MiniLM-L-6-v2}
 & 0.80 & 10 \\

 & TPE (Warm)
 & 0.3830 & 0.3702
 & \texttt{multi-qa-mpnet-base-cos-v1}
 & \texttt{recursive\_character}
 & 1500
 & \texttt{none}
 & 0.60 & 38 \\

 & IRAH (Cold)
 & 0.3818 & 0.3632
 & \texttt{multi-qa-mpnet-base-cos-v1}
 & \texttt{sentence}
 & 512
 & \texttt{none}
 & 0.67 & 50 \\

 & IRAH (Warm)
 & 0.3837 & 0.3616
 & \texttt{bge-large-en-v1.5}
 & \texttt{sentence}
 & 832
 & \texttt{none}
 & 0.69 & 50 \\
\midrule

\multirow{5}{*}{\texttt{CoSQA}}
 & Grid Search
 & 0.3406 & 0.3350
 & \texttt{bge-large-en-v1.5}
 & \texttt{sentence}
 & 256
 & \texttt{none}
 & 1.00 & 20 \\

 & TPE (Cold)
 & 0.3391 & 0.3148
 & \texttt{bge-large-en-v1.5}
 & \texttt{token}
 & 256
 & \texttt{none}
 & 1.00 & 35 \\

 & TPE (Warm)
 & 0.3414 & 0.3414
 & \texttt{bge-large-en-v1.5}
 & \texttt{token}
 & 512
 & \texttt{none}
 & 0.80 & 27 \\

 & IRAH (Cold)
 & 0.3388 & 0.3100
 & \texttt{bge-large-en-v1.5}
 & \texttt{token}
 & 256
 & \texttt{none}
 & 1.00 & 30 \\

 & IRAH (Warm)
 & 0.3427 & 0.3411
 & \texttt{bge-large-en-v1.5}
 & \texttt{token}
 & 512
 & \texttt{none}
 & 1.00 & 20 \\

\bottomrule
\end{tabular}%
}
\end{table*}

\subsection{IRAH Search Dynamics}
\label{app:2phase}

\begin{figure}[h]
\centering
\includegraphics[width=\linewidth]{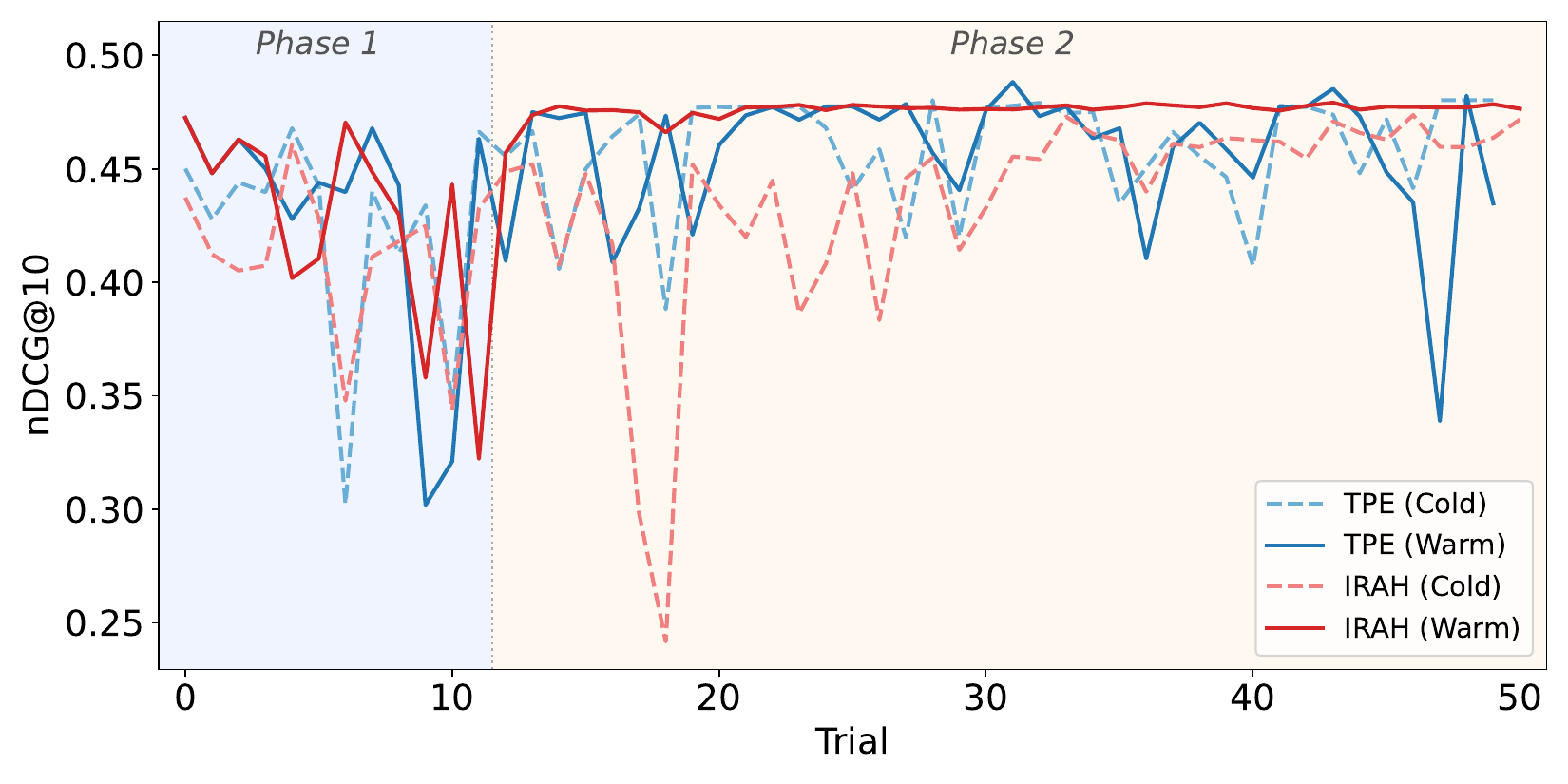}
\caption{
Trial-level optimization trajectories on \texttt{FiQA-2018} (Run~0).
The shaded regions indicate the Phase~1/Phase~2 schedule for
\mbox{IRAH (Warm)} (Trials~0--11 and 12--49, respectively).
\mbox{IRAH (Cold)} uses a longer Phase~1 schedule
(Trials~0--29), transitioning to Phase~2 at Trial~30.
TPE trajectories are shown over the same 50-trial budget for comparison.
}
\label{fig:fiqa_trajectory_phases}
\end{figure}
Figure~\ref{fig:fiqa_trajectory_phases} illustrates the trial-level search
dynamics for \texttt{FiQA-2018} Run~0. The trajectories show a qualitative
difference in the candidate configurations evaluated by the warm- and
cold-started searches. \mbox{IRAH (Cold)} exhibits substantial variation,
including a drop to approximately $0.242$ $\mathrm{nDCG}@10$ at Trial~18,
while the TPE trajectories also contain intermittent low-performing
evaluations throughout the search. In contrast, \mbox{IRAH (Warm)} enters
Phase~2 with a high-performing configuration and subsequently evaluates
configurations concentrated near $\mathrm{nDCG}@10 \approx 0.48$.

For this run, the pattern is consistent with meta-learned initialization
placing \mbox{LATS-IR} in a promising region before local Phase~2 refinement.
The resulting late-stage stability also suggests a potential direction for
future work on convergence-based early stopping when additional evaluations
produce little improvement.

\subsection{Runtime and Index Construction}
\label{app:runtime_index}

\paragraph{Grid Search Timeout Failures}
\label{app:grid_timeouts}

Across the complete Grid Search evaluation space
($108$ configurations $\times$ $4$ datasets $=432$ configuration--dataset
evaluations), 50 evaluations failed due to search-backend socket timeouts:
5 on \texttt{SciFact}, 15 on \texttt{FiQA-2018}, 17 on
\texttt{CQADupStack-STATS}, and 13 on \texttt{CoSQA}. Timed-out
configurations were treated as non-viable under the evaluation budget.


\paragraph{Selected Configurations}
\label{app:selected_configs}

Table~\ref{tab:best_configurations} reports the configuration with the
highest optimization-set $\mathrm{nDCG}@10$ for each method, together with
its optimization-set and corresponding held-out performance. For stochastic
methods, we report the selected configuration from the run achieving the
highest optimization-set score. This table is intended to characterize the
configurations found by each optimizer; aggregate held-out performance across
runs is reported in Table~\ref{tab:retrieval_results}.

The selected configurations also reveal several qualitative patterns. Within
a dataset, methods often agree on major structural components while differing
in chunking and numerical settings. For example, all methods select
\texttt{bge-large-en-v1.5} on \texttt{SciFact}, while all selected
\texttt{CoSQA} configurations use the same embedding and no reranker.
\mbox{IRAH} additionally selects off-grid chunk sizes such as 2760 on
\texttt{FiQA-2018} and 832 on \texttt{CQADupStack-STATS}, illustrating the
effect of Phase~2 local numerical refinement beyond the coarse Phase~1 grid.
Finally, optimization-set and held-out scores do not always rank runs
identically, reinforcing the use of aggregate held-out performance in
Table~\ref{tab:retrieval_results} for method comparison.

\paragraph{Index-Build Overhead}
\label{app:index-build-overhead}

Because index-defining choices such as the embedding model, chunking strategy,
and chunk size may require a distinct physical retrieval index, optimization runtime is not determined solely by the number of evaluated trials.
In contrast, changes to retrieval depth $k$, fusion weight $\alpha$, and
reranking can reuse an existing physical index in our implementation.

Table~\ref{tab:collection_overhead} therefore reports the number of unique
physical index collections required by each method. Grid Search deterministically
spans 18 collections, whereas stochastic methods report the mean and standard
deviation across three runs.

\begin{table}[h]
\centering
\footnotesize
\setlength{\tabcolsep}{3pt}
\caption{
Unique physical index collections required by each optimization method.
Stochastic methods report mean $\pm$ std.\ and range across three runs;
Grid Search is deterministic. Lowest mean collection count per dataset is shown in bold.
}
\label{tab:collection_overhead}

\resizebox{\columnwidth}{!}{%
\begin{tabular}{l l c c}
\toprule
\textbf{Dataset} & \textbf{Method} &
\textbf{Mean $\pm$ Std.} &
\textbf{Range} \\
\midrule

\multirow{5}{*}{\texttt{SciFact}}
 & Grid Search & $18.0$ & [18,18] \\
 & TPE (Cold) & $20.0 \pm 1.0$ & [19,21] \\
 & TPE (Warm) & $19.3 \pm 1.2$ & [18,20] \\
 & IRAH (Cold) & $17.3 \pm 2.1$ & [15,19] \\
 & IRAH (Warm) & $\mathbf{11.7 \pm 0.6}$ & [11,12] \\
\midrule

\multirow{5}{*}{\texttt{FiQA-2018}}
 & Grid Search & $18.0$ & [18,18] \\
 & TPE (Cold) & $20.7 \pm 1.5$ & [19,22] \\
 & TPE (Warm) & $18.0 \pm 2.0$ & [16,20] \\
 & IRAH (Cold) & $\mathbf{16.7 \pm 2.1}$ & [15,19] \\
 & IRAH (Warm) & $21.3 \pm 5.5$ & [16,27] \\
\midrule

\multirow{5}{*}{\texttt{CQADupStack-STATS}}
 & Grid Search & $18.0$ & [18,18] \\
 & TPE (Cold) & $19.7 \pm 2.3$ & [17,21] \\
 & TPE (Warm) & $20.7 \pm 1.5$ & [19,22] \\
 & IRAH (Cold) & $20.7 \pm 5.0$ & [16,26] \\
 & IRAH (Warm) & $\mathbf{15.7 \pm 3.2}$ & [12,18] \\
\midrule

\multirow{5}{*}{\texttt{CoSQA}}
 & Grid Search & $\mathbf{18.0}$ & [18,18] \\
 & TPE (Cold) & $20.3 \pm 2.3$ & [19,23] \\
 & TPE (Warm) & $19.0 \pm 1.0$ & [18,20] \\
 & IRAH (Cold) & $19.7 \pm 3.1$ & [17,23] \\
 & IRAH (Warm) & $20.7 \pm 0.6$ & [20,21] \\

\bottomrule
\end{tabular}%
}
\end{table}

The index-collection counts help contextualize the method-runtime results in
Table~\ref{tab:runtime_results}. On \texttt{SciFact}, \mbox{IRAH (Warm)}
requires only $11.7 \pm 0.6$ unique collections on average, compared with
18 for Grid Search and $17.3 \pm 2.1$ for \mbox{IRAH (Cold)}, coinciding
with a reduction in optimization time to $3.37 \pm 0.11$ hours from
$6.75$ and $5.99 \pm 1.15$ hours, respectively. On
\texttt{CQADupStack-STATS}, \mbox{IRAH (Warm)} similarly reduces both
collection count ($15.7 \pm 3.2$ versus $20.7 \pm 5.0$) and runtime
($33.28 \pm 2.53$ versus $59.92 \pm 2.75$ hours) relative to
\mbox{IRAH (Cold)}.

Collection count alone, however, does not determine runtime cost.
On \texttt{CoSQA}, \mbox{IRAH (Warm)} evaluates slightly more collections
than \mbox{IRAH (Cold)} ($20.7 \pm 0.6$ versus $19.7 \pm 3.1$), while
still reducing the reported optimization time from $5.69 \pm 2.93$ to
$2.89 \pm 0.95$ hours.

\paragraph{Runtime Diagnostics}
\label{app:runtime-decomposition}

To further examine this relationship,
Table~\ref{tab:fiqa_runtime_decomp} reports runtime diagnostics for
\texttt{FiQA-2018}, where \mbox{IRAH (Warm)} exhibits a large method-runtime overhead.

\begin{table}[h]
\centering
\footnotesize
\setlength{\tabcolsep}{3pt}
\caption{
Runtime diagnostics on \texttt{FiQA-2018}. Method runtime corresponds to
the optimization-loop wall time, while pure search time measures retrieval/evaluation time excluding explicit
collection-build cost. Values are means across three runs.
}
\label{tab:fiqa_runtime_decomp}
\resizebox{\columnwidth}{!}{%
\begin{tabular}{l c c c}
\toprule
\textbf{Method}
& \textbf{Method Runtime (h)}
& \textbf{Pure Search / Eval. (h)}
& \textbf{Collections} \\
\midrule
TPE (Warm)  & $44.71$ & $3.73$  & $18.0$ \\
IRAH (Cold) & $47.41$ & $9.71$  & $16.7$ \\
IRAH (Warm) & $74.49$ & $24.91$ & $21.3$ \\
\bottomrule
\end{tabular}}
\end{table}

On \texttt{FiQA-2018}, \mbox{IRAH (Warm)} has a method runtime of
$74.49$ hours, compared with $44.71$ hours for TPE (Warm), while its pure
search/evaluation time is also substantially larger ($24.91$ versus
$3.73$ hours). Together with the larger number of physical collections
($21.3$ versus $18.0$), these diagnostics indicate that the FiQA runtime
increase is not explained by collection count alone.

Across the three \mbox{IRAH (Warm)} runs, selected recursive-character
chunk sizes include 3120, 2950, and 2760, illustrating how Phase~2 explores
configurations outside the discrete Phase~1 grid. Thus, the higher
\texttt{FiQA-2018} runtime appears to reflect both additional index-related
work and increased search/evaluation cost rather than collection count alone.
Discretizing Phase~2 chunk-size proposals, improving index reuse, or
incorporating systems cost into the mutation policy are potential directions
for reducing this overhead.

\subsection{IRAH Agentic Token Cost}
\label{app:token_cost}

\begin{table}[!h]
\centering
\footnotesize
\setlength{\tabcolsep}{4pt}
\caption{
Agentic LLM token usage for \mbox{IRAH} across three seeds.
Values report mean total tokens per run $\pm$ standard deviation.
Reduction denotes the decrease from Cold to Warm initialization.
Grid Search and TPE do not invoke an LLM during optimization and are omitted.
}
\label{tab:token_usage}
\resizebox{\columnwidth}{!}{%
\begin{tabular}{l c c c}
\toprule
\textbf{Dataset}
& \textbf{IRAH (Cold)}
& \textbf{IRAH (Warm)}
& \textbf{Reduction} \\
\midrule

\texttt{SciFact}
& $594.1\text{k} \pm 9.8\text{k}$
& $\mathbf{530.6\text{k} \pm 6.1\text{k}}$
& $10.7\%$ \\

\texttt{FiQA-2018}
& $589.3\text{k} \pm 6.1\text{k}$
& $\mathbf{568.3\text{k} \pm 7.3\text{k}}$
& $3.6\%$ \\

\texttt{CQADupStack-STATS}
& $595.9\text{k} \pm 29.9\text{k}$
& $\mathbf{559.5\text{k} \pm 2.7\text{k}}$
& $6.1\%$ \\

\texttt{CoSQA}
& $581.1\text{k} \pm 10.4\text{k}$
& $\mathbf{565.9\text{k} \pm 6.2\text{k}}$
& $2.6\%$ \\

\midrule
Average
& $590.1\text{k}$
& $\mathbf{556.1\text{k}}$
& $\mathbf{5.8\%}$ \\

\bottomrule
\end{tabular}}
\end{table}

Table~\ref{tab:token_usage} reports the LLM token overhead associated with
reflection and mutation in \mbox{IRAH}. Warm-started IRAH uses fewer total
tokens than its cold-started counterpart on all four benchmarks, with
reductions ranging from $2.6\%$ on \texttt{CoSQA} to $10.7\%$ on
\texttt{SciFact}, and an average reduction of approximately $5.8\%$.
Because both variants use the same retrieval-evaluation budget, these differences reflect changes in agentic reasoning and context overhead rather than fewer retrieval evaluations. These reductions are consistent with the combined effects of meta-learned
initialization and the warm-start phase allocation, which alter the search
trajectory and associated LLM context overhead while preserving the same
retrieval-evaluation budget.

\newpage
\clearpage
\onecolumn

\begin{small}
\setlength{\tabcolsep}{4pt}
\begin{longtable}{@{}lllllll@{}}
\caption{The 108 pipeline configurations spanning 18 physical indexes used for Grid Search. The first 36 configurations are also used for meta-learner training.} \label{tab:appendix_all_108_configs} \\

\toprule
\textbf{Index} & \textbf{Dense Embedder} & \textbf{Chunking Strategy} & \textbf{Config ID} & \textbf{$\alpha$} & \textbf{$k$} & \textbf{Reranker Model} \\
\midrule
\endfirsthead

\multicolumn{7}{c}{\textit{(Table \ref{tab:appendix_all_108_configs} continued from previous page)}} \\
\toprule
\textbf{Index} & \textbf{Dense Embedder} & \textbf{Chunking Strategy} & \textbf{Config ID} & \textbf{$\alpha$} & \textbf{$k$} & \textbf{Reranker Model} \\
\midrule
\endhead

\midrule
\endfoot

\bottomrule
\endlastfoot

\multicolumn{7}{c}{\textbf{PART 1: Meta-Learner Training Anchors \& Grid Search Baseline (\texttt{cfg\_01}--\texttt{cfg\_36})}} \\
\midrule

\multirow{6}{*}{\textbf{1}} & \multirow{6}{*}{\texttt{bge-m3}} & \multirow{6}{*}{\texttt{token} (512, 50)} 
  & \texttt{cfg\_01} & 1.00 & 10 & \textit{none} \\
& & & \texttt{cfg\_02} & 0.67 & 30 & \textit{none} \\
& & & \texttt{cfg\_03} & 0.33 & 20 & \texttt{bge-reranker-v2-m3} \\
& & & \texttt{cfg\_04} & 0.00 & 50 & \texttt{ms-marco-MiniLM-L-6-v2} \\
& & & \texttt{cfg\_05} & 1.00 & 50 & \texttt{bge-reranker-large} \\
& & & \texttt{cfg\_06} & 0.33 & 10 & \textit{none} \\
\cmidrule{1-7}

\multirow{6}{*}{\textbf{2}} & \multirow{6}{*}{\texttt{bge-m3}} & \multirow{6}{*}{\texttt{none} (full doc)} 
  & \texttt{cfg\_07} & 1.00 & 10 & \textit{none} \\
& & & \texttt{cfg\_08} & 0.67 & 30 & \texttt{ms-marco-MiniLM-L-6-v2} \\
& & & \texttt{cfg\_09} & 0.33 & 20 & \textit{none} \\
& & & \texttt{cfg\_10} & 0.00 & 50 & \textit{none} \\
& & & \texttt{cfg\_11} & 0.67 & 10 & \texttt{bge-reranker-large} \\
& & & \texttt{cfg\_12} & 0.00 & 30 & \texttt{bge-reranker-v2-m3} \\
\cmidrule{1-7}

\multirow{6}{*}{\textbf{3}} & \multirow{6}{*}{\texttt{bge-large-v1.5}} & \multirow{6}{*}{\texttt{recursive} (3000, 300)} 
  & \texttt{cfg\_13} & 1.00 & 10 & \texttt{ms-marco-MiniLM-L-6-v2} \\
& & & \texttt{cfg\_14} & 0.67 & 30 & \texttt{bge-reranker-v2-m3} \\
& & & \texttt{cfg\_15} & 0.33 & 20 & \textit{none} \\
& & & \texttt{cfg\_16} & 0.00 & 50 & \texttt{bge-reranker-large} \\
& & & \texttt{cfg\_17} & 1.00 & 50 & \textit{none} \\
& & & \texttt{cfg\_18} & 0.33 & 10 & \texttt{bge-reranker-v2-m3} \\
\cmidrule{1-7}

\multirow{6}{*}{\textbf{4}} & \multirow{6}{*}{\texttt{bge-large-v1.5}} & \multirow{6}{*}{\texttt{none} (full doc)} 
  & \texttt{cfg\_19} & 1.00 & 20 & \texttt{bge-reranker-large} \\
& & & \texttt{cfg\_20} & 0.67 & 30 & \textit{none} \\
& & & \texttt{cfg\_21} & 0.33 & 50 & \texttt{bge-reranker-v2-m3} \\
& & & \texttt{cfg\_22} & 0.00 & 10 & \textit{none} \\
& & & \texttt{cfg\_23} & 0.67 & 10 & \texttt{ms-marco-MiniLM-L-6-v2} \\
& & & \texttt{cfg\_24} & 0.00 & 50 & \texttt{bge-reranker-large} \\
\cmidrule{1-7}

\multirow{6}{*}{\textbf{5}} & \multirow{6}{*}{\texttt{mpnet-base}} & \multirow{6}{*}{\texttt{recursive} (3000, 300)} 
  & \texttt{cfg\_25} & 1.00 & 10 & \texttt{ms-marco-MiniLM-L-6-v2} \\
& & & \texttt{cfg\_26} & 0.67 & 30 & \textit{none} \\
& & & \texttt{cfg\_27} & 0.33 & 20 & \texttt{bge-reranker-v2-m3} \\
& & & \texttt{cfg\_28} & 0.00 & 50 & \texttt{bge-reranker-large} \\
& & & \texttt{cfg\_29} & 1.00 & 30 & \textit{none} \\
& & & \texttt{cfg\_30} & 0.33 & 50 & \texttt{ms-marco-MiniLM-L-6-v2} \\
\cmidrule{1-7}

\multirow{6}{*}{\textbf{6}} & \multirow{6}{*}{\texttt{mpnet-base}} & \multirow{6}{*}{\texttt{sentence} (512, 50)} 
  & \texttt{cfg\_31} & 1.00 & 50 & \textit{none} \\
& & & \texttt{cfg\_32} & 0.67 & 10 & \texttt{bge-reranker-large} \\
& & & \texttt{cfg\_33} & 0.33 & 20 & \textit{none} \\
& & & \texttt{cfg\_34} & 0.00 & 30 & \texttt{bge-reranker-v2-m3} \\
& & & \texttt{cfg\_35} & 0.67 & 50 & \texttt{ms-marco-MiniLM-L-6-v2} \\
& & & \texttt{cfg\_36} & 0.00 & 10 & \textit{none} \\

\midrule[1.5pt]
\multicolumn{7}{c}{\textbf{PART 2: Expanded Grid Search Baseline Only (\texttt{cfg\_37}--\texttt{cfg\_108})}} \\
\midrule

\multirow{6}{*}{\textbf{7}} & \multirow{6}{*}{\texttt{bge-m3}} & \multirow{6}{*}{\texttt{token} (256, 25)} 
  & \texttt{cfg\_37} & 1.00 & 10 & \textit{none} \\
& & & \texttt{cfg\_38} & 0.67 & 20 & \texttt{ms-marco-MiniLM-L-6-v2} \\
& & & \texttt{cfg\_39} & 0.33 & 30 & \textit{none} \\
& & & \texttt{cfg\_40} & 0.00 & 50 & \texttt{bge-reranker-v2-m3} \\
& & & \texttt{cfg\_41} & 0.67 & 50 & \texttt{bge-reranker-large} \\
& & & \texttt{cfg\_42} & 1.00 & 30 & \textit{none} \\
\cmidrule{1-7}

\multirow{6}{*}{\textbf{8}} & \multirow{6}{*}{\texttt{bge-m3}} & \multirow{6}{*}{\texttt{token} (512, 100)} 
  & \texttt{cfg\_43} & 1.00 & 20 & \texttt{bge-reranker-v2-m3} \\
& & & \texttt{cfg\_44} & 0.67 & 10 & \textit{none} \\
& & & \texttt{cfg\_45} & 0.33 & 50 & \texttt{ms-marco-MiniLM-L-6-v2} \\
& & & \texttt{cfg\_46} & 0.00 & 30 & \texttt{bge-reranker-large} \\
& & & \texttt{cfg\_47} & 1.00 & 50 & \textit{none} \\
& & & \texttt{cfg\_48} & 0.33 & 20 & \textit{none} \\
\cmidrule{1-7}

\multirow{6}{*}{\textbf{9}} & \multirow{6}{*}{\texttt{bge-large-v1.5}} & \multirow{6}{*}{\texttt{token} (512, 50)} 
  & \texttt{cfg\_49} & 1.00 & 10 & \textit{none} \\
& & & \texttt{cfg\_50} & 0.67 & 30 & \texttt{bge-reranker-large} \\
& & & \texttt{cfg\_51} & 0.33 & 20 & \texttt{ms-marco-MiniLM-L-6-v2} \\
& & & \texttt{cfg\_52} & 0.00 & 50 & \textit{none} \\
& & & \texttt{cfg\_53} & 1.00 & 50 & \texttt{bge-reranker-v2-m3} \\
& & & \texttt{cfg\_54} & 0.67 & 10 & \textit{none} \\
\cmidrule{1-7}

\multirow{6}{*}{\textbf{10}} & \multirow{6}{*}{\texttt{bge-large-v1.5}} & \multirow{6}{*}{\texttt{sentence} (256, 25)} 
  & \texttt{cfg\_55} & 1.00 & 20 & \textit{none} \\
& & & \texttt{cfg\_56} & 0.67 & 10 & \texttt{ms-marco-MiniLM-L-6-v2} \\
& & & \texttt{cfg\_57} & 0.33 & 50 & \texttt{bge-reranker-v2-m3} \\
& & & \texttt{cfg\_58} & 0.00 & 30 & \texttt{bge-reranker-large} \\
& & & \texttt{cfg\_59} & 1.00 & 50 & \textit{none} \\
& & & \texttt{cfg\_60} & 0.33 & 10 & \textit{none} \\
\cmidrule{1-7}

\multirow{6}{*}{\textbf{11}} & \multirow{6}{*}{\texttt{bge-large-v1.5}} & \multirow{6}{*}{\texttt{sentence} (512, 50)} 
  & \texttt{cfg\_61} & 1.00 & 10 & \texttt{bge-reranker-v2-m3} \\
& & & \texttt{cfg\_62} & 0.67 & 30 & \textit{none} \\
& & & \texttt{cfg\_63} & 0.33 & 20 & \texttt{bge-reranker-large} \\
& & & \texttt{cfg\_64} & 0.00 & 50 & \texttt{ms-marco-MiniLM-L-6-v2} \\
& & & \texttt{cfg\_65} & 1.00 & 30 & \textit{none} \\
& & & \texttt{cfg\_66} & 0.67 & 50 & \textit{none} \\
\cmidrule{1-7}

\multirow{6}{*}{\textbf{12}} & \multirow{6}{*}{\texttt{bge-large-v1.5}} & \multirow{6}{*}{\texttt{sentence} (512, 100)} 
  & \texttt{cfg\_67} & 1.00 & 20 & \textit{none} \\
& & & \texttt{cfg\_68} & 0.67 & 50 & \texttt{ms-marco-MiniLM-L-6-v2} \\
& & & \texttt{cfg\_69} & 0.33 & 10 & \texttt{bge-reranker-large} \\
& & & \texttt{cfg\_70} & 0.00 & 30 & \texttt{bge-reranker-v2-m3} \\
& & & \texttt{cfg\_71} & 0.67 & 20 & \textit{none} \\
& & & \texttt{cfg\_72} & 1.00 & 50 & \textit{none} \\
\cmidrule{1-7}

\multirow{6}{*}{\textbf{13}} & \multirow{6}{*}{\texttt{mpnet-base}} & \multirow{6}{*}{\texttt{token} (256, 25)} 
  & \texttt{cfg\_73} & 1.00 & 10 & \textit{none} \\
& & & \texttt{cfg\_74} & 0.67 & 30 & \texttt{bge-reranker-large} \\
& & & \texttt{cfg\_75} & 0.33 & 20 & \texttt{ms-marco-MiniLM-L-6-v2} \\
& & & \texttt{cfg\_76} & 0.00 & 50 & \textit{none} \\
& & & \texttt{cfg\_77} & 1.00 & 50 & \texttt{bge-reranker-v2-m3} \\
& & & \texttt{cfg\_78} & 0.33 & 10 & \textit{none} \\
\cmidrule{1-7}

\multirow{6}{*}{\textbf{14}} & \multirow{6}{*}{\texttt{mpnet-base}} & \multirow{6}{*}{\texttt{token} (512, 50)} 
  & \texttt{cfg\_79} & 1.00 & 20 & \texttt{bge-reranker-v2-m3} \\
& & & \texttt{cfg\_80} & 0.67 & 10 & \textit{none} \\
& & & \texttt{cfg\_81} & 0.33 & 50 & \texttt{ms-marco-MiniLM-L-6-v2} \\
& & & \texttt{cfg\_82} & 0.00 & 30 & \texttt{bge-reranker-large} \\
& & & \texttt{cfg\_83} & 0.67 & 50 & \textit{none} \\
& & & \texttt{cfg\_84} & 1.00 & 30 & \textit{none} \\
\cmidrule{1-7}

\multirow{6}{*}{\textbf{15}} & \multirow{6}{*}{\texttt{mpnet-base}} & \multirow{6}{*}{\texttt{sentence} (256, 25)} 
  & \texttt{cfg\_85} & 1.00 & 10 & \textit{none} \\
& & & \texttt{cfg\_86} & 0.67 & 20 & \texttt{ms-marco-MiniLM-L-6-v2} \\
& & & \texttt{cfg\_87} & 0.33 & 30 & \texttt{bge-reranker-v2-m3} \\
& & & \texttt{cfg\_88} & 0.00 & 50 & \texttt{bge-reranker-large} \\
& & & \texttt{cfg\_89} & 1.00 & 50 & \textit{none} \\
& & & \texttt{cfg\_90} & 0.67 & 10 & \textit{none} \\
\cmidrule{1-7}

\multirow{6}{*}{\textbf{16}} & \multirow{6}{*}{\texttt{mpnet-base}} & \multirow{6}{*}{\texttt{recursive} (1500, 150)} 
  & \texttt{cfg\_91} & 1.00 & 20 & \texttt{bge-reranker-large} \\
& & & \texttt{cfg\_92} & 0.67 & 30 & \textit{none} \\
& & & \texttt{cfg\_93} & 0.33 & 10 & \texttt{ms-marco-MiniLM-L-6-v2} \\
& & & \texttt{cfg\_94} & 0.00 & 50 & \texttt{bge-reranker-v2-m3} \\
& & & \texttt{cfg\_95} & 1.00 & 50 & \textit{none} \\
& & & \texttt{cfg\_96} & 0.33 & 20 & \textit{none} \\
\cmidrule{1-7}

\multirow{6}{*}{\textbf{17}} & \multirow{6}{*}{\texttt{bge-m3}} & \multirow{6}{*}{\texttt{recursive} (1500, 150)} 
  & \texttt{cfg\_97} & 1.00 & 10 & \textit{none} \\
& & & \texttt{cfg\_98} & 0.67 & 20 & \texttt{bge-reranker-v2-m3} \\
& & & \texttt{cfg\_99} & 0.33 & 50 & \texttt{ms-marco-MiniLM-L-6-v2} \\
& & & \texttt{cfg\_100} & 0.00 & 30 & \texttt{bge-reranker-large} \\
& & & \texttt{cfg\_101} & 0.67 & 50 & \textit{none} \\
& & & \texttt{cfg\_102} & 1.00 & 30 & \textit{none} \\
\cmidrule{1-7}

\multirow{6}{*}{\textbf{18}} & \multirow{6}{*}{\texttt{bge-m3}} & \multirow{6}{*}{\texttt{recursive} (3000, 600)} 
  & \texttt{cfg\_103} & 1.00 & 20 & \texttt{bge-reranker-large} \\
& & & \texttt{cfg\_104} & 0.67 & 10 & \textit{none} \\
& & & \texttt{cfg\_105} & 0.33 & 50 & \texttt{bge-reranker-v2-m3} \\
& & & \texttt{cfg\_106} & 0.00 & 30 & \texttt{ms-marco-MiniLM-L-6-v2} \\
& & & \texttt{cfg\_107} & 1.00 & 50 & \textit{none} \\
& & & \texttt{cfg\_108} & 0.33 & 20 & \textit{none} \\
\end{longtable}
\end{small}

\end{document}